\documentclass[usenatbib]{mn2e}
\usepackage[utf8]{inputenc}
\usepackage{natbib,epsfig,cite,mystyle,amsmath,amssymb,graphicx}
\usepackage{array, xcolor, caption}
\usepackage{hyperref}
\hypersetup{colorlinks,allcolors=black}

\pdfoutput=1  % togliere il commento se le figure sono pdf, lasciarlo se sono eps
\usepackage{bm}
\usepackage{graphicx,color}
\usepackage{latexsym,amsmath,amssymb,graphicx,booktabs}
\usepackage{placeins}
\usepackage{subfigure}
\usepackage{mathtools}
\usepackage[normalem]{ulem}

\definecolor{MyBlue}{rgb}{0.15,0.15,0.70}
\definecolor{Dgreen}{rgb}{0,0.7,0.0}

\hypersetup{
colorlinks=true,
citecolor=MyBlue,
linkcolor=MyBlue,
urlcolor=MyBlue
}

\usepackage{amssymb}
\usepackage{amsmath}
\usepackage{amsfonts}
\usepackage{upgreek}
\usepackage{latexsym}
\usepackage{appendix}
\usepackage{dsfont}
\usepackage{mathabx}

\newcommand\spart{\;\raise1.0pt\hbox{/}\hskip-6pt\partial}
\newcommand\spartb{\;\overline{\raise1.0pt\hbox{/}\hskip-6pt\partial}}

\newcommand{\be}{\begin{equation}}
\newcommand{\ee}{\end{equation}}

\newcommand{\data}{\boldsymbol{d}}
\newcommand{\dth}{\ensuremath{\boldsymbol d_\text{th}}}
\newcommand{\pdetl}{\ensuremath{P_{\rm det}(\vec{\lambda})}}
\newcommand{\sigmaplussigma}{\ensuremath{(\sigma^2+\Sigma^2)}}

\begin{document}
\title[A Fisher matrix for GW population inference]{A Fisher matrix for gravitational-wave population inference}
\author[J.R.~Gair, A.~Antonelli, R.~Barbieri]{Jonathan R.~Gair$^{1}$\thanks{jonathan.gair@aei.mpg.de}, Andrea Antonelli$^{2}$\thanks{aantone3@jh.edu}, Riccardo Barbieri$^{1}$\thanks{riccardo.barbieri@aei.mpg.de} \\
$^{1}$Max Planck Institute for Gravitational Physics (Albert Einstein Institute), Am M\"{u}hlenberg 1, Potsdam-Golm, 14476, Germany \\
% $^{2}$School of Mathematics, University of Edinburgh, James Clerk Maxwell Building, Peter Guthrie Tait Road, Edinburgh EH9 3FD, UK \\
$^{2}$Department of Physics and Astronomy, Johns Hopkins University,
3400 N. Charles Street, Baltimore, Maryland, 21218, USA
}

\pagerange{\pageref{firstpage}--\pageref{lastpage}} \pubyear{2021}
\maketitle
\label{firstpage}

\begin{abstract}

We derive a Fisher matrix for the parameters characterising a population of gravitational-wave events. This provides a guide to the precision with which population parameters can be estimated with multiple observations, which becomes increasingly accurate as the number of events and the signal-to-noise ratio of the sampled events increases. The formalism takes into account individual event measurement uncertainties and selection effects, and can be applied to arbitrary population models.
We illustrate the framework with two examples: an analytical calculation of the Fisher matrix for the mean and variance of a Gaussian model describing a population affected by selection effects, and an estimation of the precision with which the slope of a power law distribution of supermassive black-hole masses can be measured using extreme-mass-ratio inspiral observations.  We compare the Fisher predictions to results from Monte Carlo analyses, finding very good agreement.
\end{abstract}


%%%%%%%%%%%%%%%%%%%%%%%%%%%%%%%%%%%%%%%%%%%%%%%%%%
\section{Introduction} 

Population analyses aim at inferring the parameters that describe the distribution of the properties of a set of observed events drawn from a common population. In the context of gravitational-wave (GW) astrophysics, such analyses have been carried out for the 90 coalescing compact-object binaries that have so far been observed by ground-based gravitational wave detectors, and reported in the third gravitational wave transient catalogue, GWTC-3~\citep{LIGOScientific:2021usb,LIGOScientific:2021djp,LIGOScientific:2021psn}. {Together with simulation-based studies \citep{Taylor:2018iat}, these population analyses aimed at understanding the astrophysical processes that lead to the formation of the binaries \citep{LIGOScientific:2018mvr, Rodriguez:2020viw}, their evolution  \citep{Fishbach:2021yvy, Mould:2022xeu} and at measuring the current parameters describing their population \citep{Vitale:2022pmu}. Furthermore,} population analysis are also used to constrain cosmic expansion history by estimating parameters like the Hubble constant \citep{Mastrogiovanni:2021wsd,LIGOScientific:2021aug,Mancarella:2021ecn,Mukherjee:2022afz}.

Given a set of observed events, the usual approach to estimate distribution parameters is to complete a Bayesian hierarchical analysis using techniques such as Markov Chain Monte Carlo (MCMC). While these are the most reliable way to obtain posterior samples from actual data, they are typically computationally expensive and so it can become impractical to use these approaches to make forecasts for future observations that include surveys over parameter space. However, such surveys are crucial for scoping out the science cases of future detectors, such as the Einstein Telescope \citep{Punturo:2010zz}, Cosmic Explorer \citep{Reitze:2019iox} and the spaceborne LISA mission \citep{Audley:2017drz}, all of which are expected to detect thousands of sources from multiple populations. For explorations of this nature, one can trade off accuracy in the estimates of parameter-measurement precision for computational speed by using approximations that are valid in the limit of high signal-to-noise ratio (SNR). %In turn, such precision studies inform the development of future detectors. 
In the context of \emph{source} parameters for \emph{individual} signals, the Fisher matrix is commonly used to cheaply assess the measurement precision of a parameter \citep{Vallisneri:2007ev}.
Within the linear-signal approximation, valid for high SNR sources, the inverse of the Fisher matrix is an approximation to the covariance matrix and therefore the width of the likelihood function. Under the assumption of flat priors, it also approximates the shape of the Bayesian posterior probability distribution we would expect to obtain in an MCMC analyses. For a parameter set $\vec \lambda$, the Fisher matrix can be written in general terms as the expectation value over the data generating process of derivatives of the log-likelihood $p(\mathbf{d}| \vec\lambda)$,
\be
 (\Gamma_{\lambda})_{ij} = \mathbb{E}\left[ - \frac{\partial^2 \ln p(\mathbf{d}|\vec\lambda)}{\partial \lambda^i \partial \lambda^j} \right].
 \label{eq:popfishgen}
 \ee
While this provides a guide to measurement uncertainties for individual events, the Fisher matrix does not directly provide an indication of how well the properties of the population can be inferred when those events are subsequently combined in a hierarchical model. In this paper we address this shortcoming by deriving a Fisher Matrix for the population parameters assuming Gaussian noise and using the likelihood for population inference in the presence of selection effects from \citep{Mandel:2018mve}. The expression we obtain is valid for high SNRs and small biases in the individual events' parameters.
We illustrate our formalism with two examples. First, we consider a ``Gaussian-Gaussian'' case, in which both noise and the data generation process are normally distributed, and check our expressions against the direct calculation of the Fisher matrix as an expectation value over data realizations. We also perform MCMC analyses with and without selection effects, as a cross-check to verify our results.
Secondly, we consider the more astrophysically relevant case of a power-law distributed population (while still assuming Gaussian noise) and again validate our results against MCMC analyses with and without selection effects. We generally find an excellent agreement between the Fisher and MCMC estimates, while confirming results in \citep{Gair:2010yu} for the latter scenario.

The paper is organized as follows. In Section \ref{sec:derivation_FM} we describe our population Fisher Matrix formalism, highlighting the main assumptions and steps to obtain the result. %model and noise generic formula, leaving the details of the calculation 
A derivation of corrections to this formula and their scalings is found in Appendix~\ref{app:popFMcorrect}. In Section \ref{sec:gaussian-gaussian}, we consider the Gaussian population model, checking our formula against a direct calculation of the Fisher Matrix and an MCMC analysis. In section \ref{sec:EMRI-model}, we consider the case of inference of a power-law massive black hole mass distribution using extreme-mass-ratio inspiral (EMRI) observations, once again comparing the result against MCMC. Finally, in section \ref{sec:conclusions} we discuss our results and prospects for future work. The framework we develop here could be applied in a wide variety of contexts. The focus on gravitational wave detectors and the choice of the examples provided here are driven purely by the authors' areas of expertise. The results can be fully reproduced with codes made publicly available at \url{https://github.com/aantonelli94/PopFisher}.

%%%%%%%%%%%%%%%%%%%%%%%%%%%%%%%%%%%%%%%%%%%%%%%%%%
\section{The Fisher matrix for population distributions} 
%%%%%%%%%%%%%%%%%%%%%%%%%%%%%%%%%%%%%%%%%%%%%%%%%%
\label{sec:derivation_FM}

% \jg{I have now done this section and included all of the parts that I think are needed. Please do read through it and make sure it is sufficiently clear. Also let me know if you think more details are needed in the final part of the derivation.} \andrea{A few more details after Eq. 20 would be nice.} \jg{Added a few more details. Please review.}
The standard model used to represent the data stream, $\mathbf{d}$, of a gravitational wave detector is as a linear combination of a signal, $\mathbf{h}(\vec\theta)$, dependent on some parameters $\vec\theta$, and noise, $\mathbf{n}$, that is usually assumed to be a realisation of a stationary and Gaussian stochastic process described by a power spectral density $S_h(f)$,
\be
\mathbf{d} = \mathbf{h}(\vec\theta) + \mathbf{n}, \qquad \langle \tilde{n}^*(f) \tilde{n}(f')\rangle = S_h(f) \delta(f-f').
\ee
In this model the likelihood is
\begin{align}
p(\mathbf{d} | \vec\theta) &\propto \exp\left[ -\frac{1}{2} \left( \mathbf{d} - \mathbf{h}(\vec\theta) | \mathbf{d} - \mathbf{h}(\vec\theta)\right) \right], \nonumber \\
\mbox{where } (\mathbf{a}|\mathbf{b}) &= 4\mbox{Re} \,\int_{0}^\infty \frac{\tilde{a}^*(f) \tilde{b}(f)}{S_h(f)} \, {\rm d}f. \label{eq:innprod}
\end{align}
To understand the precision with which gravitational wave observations can determine the parameters of a source, it is common to compute the Fisher information matrix, defined by
\be
 (\Gamma_{\theta})_{ij} = \mathbb{E}\left[ \frac{\partial \ln p(\mathbf{d}|\vec\theta)}{\partial \theta^i} \frac{\partial \ln p(\mathbf{d}|\vec\theta)}{\partial \theta^j} \right] \label{eq:eventfishgen},
 %\equiv \mathbb{E}\left[ - \frac{\partial^2 \ln p(\mathbf{d}|\vec\theta)}{\partial \theta^i \partial \thetaa^j} \right]
 \ee
where the expectation value is taken over realizations of the data drawn from the data generating process, $\mathbf{d}$. For the gravitational wave detector likelihood in Eq.~\eqref{eq:innprod}, the Fisher information matrix can be seen to reduce to
\be
(\Gamma_{\theta})_{ij} = \left( \frac{\partial \mathbf{h}}{\partial\theta^i}\bigg| \frac{\partial \mathbf{h}}{\partial\theta^j} \right), \label{eq:sourceFM}
\ee
where we are using the inner product introduced in Eq.~\eqref{eq:innprod}. 
The Fisher matrix provides a leading order approximation to the shape of the likelihood and hence also the Bayesian posterior when using priors that are approximately flat over the support of the likelihood. It becomes an increasingly good guide to the precision of parameter estimation as the SNR with which the source is observed increases.

In population inference, we are no longer primarily interested in the parameters of the individual events, but in the parameters that characterise the population from which the individual events are drawn. We assume that we have some population model, $p(\vec\theta|\vec\lambda)$, that describes the probability distribution of the parameters, $\vec\theta$, of individual events drawn randomly from a population characterised by parameters, $\vec\lambda$. We want to infer the parameters of the population by combining the information from many observed events. For a given choice of population parameters, the distribution of observed datasets is characterised by
\begin{align}
p(\mathbf{d}|\vec\lambda) &= \frac{p_{\rm full}(\mathbf{d}|\vec\lambda)}{P_{\rm det}(\vec\lambda)} \label{eq:likewithsel}\\
\mbox{where } p_{\rm full}(\mathbf{d}|\vec\lambda) &=
\int p(\mathbf{d}|\vec\theta) p(\vec\theta | \vec\lambda) \, {\rm d} \vec\theta \nonumber \\
P_{\rm det}(\vec\lambda) &= \int P_{\rm det}(\vec\theta) p(\vec\theta | \vec\lambda) \, {\rm d} \vec\theta \nonumber \\
P_{\rm det}(\vec\theta) &= \int_{\mathbf{d} > {\rm thresh}} p(\mathbf{d}|\vec\theta) \, {\rm d}\mathbf{d}.
\label{eq:poplike}
\end{align}
{Here and elsewhere we will use lower-case $p(x)$ to denote probability density functions, which have units of $1/x$, and upper-case $P(x)$ to denote cumulative density functions, which are dimensionless.} This expression accounts for the fact that not all events that occur in the Universe are detected. Detection is a property of the observed data, $\mathbf{d}$, and the last integral is over all data sets that would pass the threshold to be counted as a detected event and hence included in the population inference. The normalisation term, $P_{\rm det}(\vec\lambda)$, depends only on the population parameters and represents the fraction of events in the Universe that are detectable. We refer the reader to~\citep{Mandel:2018mve} for further details. This form of the likelihood assumes that the number of events observed in a fixed time period does not convey any information about the population parameters. However, the precision with which the population parameters are estimated asymptotically is independent of that assumption. This is discussed in more detail in Appendix~\ref{app:rates}.

Equation~\eqref{eq:popfishgen} is the equivalent of Eq.~\eqref{eq:eventfishgen} for this population likelihood, and so it should give a guide to the precision with which the population parameters can be measured. Note that the two forms of the expression are slightly different, but it is straightforward to show that the two results are equivalent by integrating by parts and using conservation of probability. 
% \jg{Are we bothered by this change in form of the expression?} \andrea{No. If the referee asks we can be more specific.} 
This will be a good guide for a ``high signal-to-noise ratio'', which for populations means a large number of observed events. The fact that Eq.~\eqref{eq:popfishgen} is a good approximation to the precision of population inference can be seen as follows. In a general population inference problem, we have observed a set of events, indexed by $i$, with corresponding datasets $\{\mathbf{d}_i\}$. The posterior distribution on the population parameters from this set of events can be found from Bayes' theorem and takes the form
 \be
 p(\vec{\lambda} | \left\{\mathbf{d}_i\right\} ) \propto \pi(\vec\lambda) \prod_{i=1}^n p(\mathbf{d}_i | \vec\lambda)
 \label{eq:jointlike}
 \ee
 where $n$ is the total number of events observed, $\pi(\vec\lambda)$ is the prior on the population parameters and $p(\mathbf{d}_i | \vec\lambda)$ is the likelihood of the population parameters $\vec\lambda$ for dataset $\mathbf{d}_i$. The log-posterior is
 \be
 \ln p(\vec{\lambda} | \left\{\mathbf{d}_i\right\} ) \propto \ln\pi(\vec\lambda) + \sum_{i=1}^n \ln p(\mathbf{d}_i | \vec\lambda).
 \ee
The latter quantity is a sum of independent random variables (assuming that all observations are independent). In the limit that $n \rightarrow \infty$ we can use the central limit theorem to deduce
\be 
 \frac{1}{n} \sum_{i=1}^n \ln p(\mathbf{d}_i | \vec\lambda) \sim N\left( \mu(\vec\lambda | \vec\lambda_t), \frac{\sigma^2(\vec\lambda  | \vec\lambda_t)}{n}\right)
 \ee
 where
 \be
 \mu(\vec\lambda | \vec\lambda_t) = \mathbb{E} \left[\ln p(\mathbf{d} | \vec\lambda)\right], \quad \sigma^2(\vec\lambda | \vec\lambda_t) =  \mathbb{E} \left[(\ln p(\mathbf{d} | \vec\lambda) - \mu)^2\right]
 \ee
 and the expectation value is taken over the data generating process, which we assume to be consistent with the likelihood we are using, evaluated for the true values of the population parameters $\vec\lambda_t$. %In other words
 %\be
 %\mathbb{E}_t\left[X(\mathbf{d})\right] \equiv \int X(\mathbf{d}) p_t(\mathbf{d}|\vec\lambda_t) {\rm d}\mathbf{d} .
 %\ee
 Since
 \begin{align}%\label{eq: dm_dlambda}
     \frac{\partial \mu}{\partial \lambda^i} &= \int p(\mathbf{d} | \vec\lambda_t) \frac{1}{p(\mathbf{d} |\vec\lambda)} \frac{\partial p(\mathbf{d}|\vec\lambda)}{\partial \lambda^i} \, {\rm d}\mathbf{d} \nonumber \\
     \Rightarrow \quad \frac{\partial \mu}{\partial \lambda^i}_{\vec\lambda=\vec\lambda_t} &= \int \frac{\partial p(\mathbf{d}|\vec\lambda)}{\partial \lambda^i} \, {\rm d}\mathbf{d} = \frac{\partial }{\partial \lambda^i}\,\int p(\mathbf{d}|\vec\lambda) \, {\rm d}\mathbf{d} \nonumber \\
     &= \frac{\partial }{\partial \lambda^i}(1) = 0, \label{eq:unbiasproof}
 \end{align}
 we deduce that the population likelihood is peaked at the true parameters asymptotically (this is not true on average for a finite number of observations, as discussed in Appendix~\ref{app:popFMcorrect}). Note that this happens by virtue of the assumed consistency between the likelihood and the data-generating process, and would not be the case if the likelihood was only an approximation to the true population. 
% We denote by $\vec\Lambda_{\rm m}(\vec\lambda_t)$ the location of the maximum of $\mu(\vec\lambda | \vec\lambda_t)$. 
As $n \rightarrow \infty$ the log-posterior converges to the function $n \mu(\vec\lambda)$, and so the posterior becomes increasingly concentrated around $\vec\lambda_t$. Expanding the function $\mu(\vec\lambda | \vec\lambda_t)$ near $\vec\lambda_t$ we have
 \begin{align}
   \mu(\vec\lambda | \vec\lambda_t)& = \mu\left(\vec\lambda_t | \vec\lambda_t\right) \nonumber\\
   +& \frac{1}{2} \left( \frac{{\rm d}^2 \mu}{{\rm d} \lambda^i {\rm d}\lambda^j}\right)_{
   \vec\lambda_t%\vec\Lambda_{\rm m}(\vec\lambda_t)
   } (\lambda^i - \lambda^i_t) (\lambda^j - \lambda^j_t) + \cdots .
 \end{align}
 We deduce that the asymptotic covariance matrix is $\Gamma_\lambda^{-1}/n$, where
  \be
  (\Gamma_{\lambda})_{ij} = -\left( \frac{{\rm d}^2 \mu}{{\rm d} \lambda^i {\rm d}\lambda^j}\right)_{
  \vec\lambda_t%\Lambda_{\rm m}(\vec\lambda_t)
  } = \mathbb{E}\left[ - \frac{\partial^2 \ln p(\mathbf{d}|\vec\lambda)}{\partial \lambda^i \partial \lambda^j} \right]. \label{eq:asymptotic_cov}
  \ee
In the last equality, we use Eq. \eqref{eq:unbiasproof}.
This justifies the use of Eq.~\eqref{eq:popfishgen} to characterise the precision of parameter estimation in the limit $n\rightarrow\infty$. It becomes increasingly reliable as $n\rightarrow \infty$, as corrections to this formula scale like $n^{-\frac{1}{2}}$ relative to leading order. This is justified in more detail in Appendix~\ref{app:popFMcorrect}.

This result can be evaluated at various levels of approximation. The full asymptotic posterior is described by the function $\mu(\vec\lambda|\vec\lambda_t)$, which can be evaluated through Monte Carlo integration. This is computationally expensive as it requires evaluation over different choices of $\vec\lambda$ and $\vec\lambda_t$. The next level of approximation is to evaluate Eq.~\eqref{eq:popfishgen} directly. This makes a linear signal approximation in the population parameters, but no approximation to the evaluation of $p(\mathbf{d}|\vec\lambda)$. This is less complex because evaluation is only needed in the vicinity of $\vec\lambda_t$. A final level of approximation is to simplify $p(\mathbf{d}|\vec\lambda)$ by using the linear signal approximation for the individual event parameters as well. This is the approach we will now describe.

% These results are valid asymptotically, but to be useful in practice we need to know how many observations are sufficient for these results to be a good approximation. To do this, we must derive the first corrections to these results for finite $n$.

% We now take the linear signal approximation one step further and suppose that the posterior for each individual event only has support in a small enough vicinity around the true parameters that we can use the linear signal approximation.
We consider a single observation of a source with parameters 
$\vec\theta_0$, and data $\mathbf{d} = \mathbf{h}(\vec\theta_0) + \mathbf{n}$. 
% \andrea{$\mathbf{h}_t$ is redundant. Waveforms are considered exact in our paper. We define $\mathbf{d}$ with $\mathbf{h}(\vec\theta_0)$ here, but use $\mathbf{h}(\vec\theta)$ in the likelihood below.}
% \jg{Deleted the t subscript. The difference between $\theta_0$ and $\theta$ is between the true value in the data, and the value at which we are evaluating the likelihood, so these should be different.}
Taking the expectation value over the true data distribution then reduces to taking the expectation value over the distribution of the noise $\mathbf{n}$ and the distribution of the parameters $\vec\theta_0$, which is $p(\vec\theta_0 | \vec\lambda_t)$. Under the linear signal approximation we expand 
\be
\mathbf{h}(\vec\theta) = \mathbf{h}(\vec\theta_0) + \frac{\partial \mathbf{h}}{\partial \theta^i} \Delta \theta^i
\ee
%and keep only the leading terms in $\Delta\vec\theta$.
where $\Delta\theta^i = \theta^i - \theta_0^i$. The gravitational wave likelihood can then be written
\begin{align}
\tilde{p}(\{ \mathbf{d} \} | \vec\theta) &\propto \exp\left[ -\frac{1}{2} (\mathbf{d}-\mathbf{h}(\vec\theta) | \mathbf{d}-\mathbf{h}(\vec\theta))\right] \nonumber \\
&\approx\exp\left[-\frac{1}{2}\left( \mathbf{n} | \mathbf{n}\right) + N_i \Delta \theta^i -\frac{1}{2} (\Gamma_\theta)_{ij} \Delta\theta^i \Delta\theta^j \right] \nonumber \\
\mbox{where } N_i &=  \left( \frac{\partial \mathbf{h}}{\partial\theta^i} \bigg| \mathbf{n} \right) \label{eq:LSA_individ}
\end{align}
and $(\Gamma_\theta)_{ij}$ is the single source Fisher matrix defined in Eq.~\eqref{eq:sourceFM}. This is to be evaluated at $\theta_0$ and therefore has a dependence on those parameters. We similarly expand the source prior term
% \andrea{Are the expressions below intended to be evaluated at $\vec\lambda_t$?} \jg{No, at $\theta_0$, have clarified.}
 \begin{align}\label{eq:lnp_P_H}
 \ln p(\vec\theta | \vec\lambda)  &=  \ln p(\vec\theta_0 | \vec\lambda) + P_i \Delta\theta^i - \frac{1}{2} H_{ij} \Delta \theta^i \Delta \theta^j + \cdots \nonumber \\
  \mbox{where } P_i &=  \frac{\partial \ln p(\vec\theta|\vec\lambda)}{\partial \theta^i}, \qquad H_{ij} = -\frac{\partial^2 \ln p(\vec\theta|\vec\lambda)}{\partial \theta^i \partial \theta^j},
\end{align}
in which the derivatives are evaluated at the parameter space point $\theta_0$. Substituting the preceding two expressions into Eq.~\ref{eq:poplike} and integrating over $\vec\theta$, which is equivalent to integrating over $\Delta\vec\theta$ in the linear signal approximation, we obtain
% Combining this expression with the likelihood expansion and integrating over $\vec\theta$, which is equivalent in this approximation to integrating over $\Delta\vec\theta$, we obtain
 \begin{align}%\label{pdlambda}
 \tilde{p}(\{ \mathbf{d} \} | \vec\lambda) &\approx \frac{\exp[-(\mathbf{n}|\mathbf{n})/2]}{P_{\rm det}(\vec\lambda)} \int {\rm d}\Delta\vec\theta \left[ p(\vec\theta_0|\vec\lambda) \right. \nonumber \\
 &\hspace{0.5cm}
 \exp\left\{ -\frac{1}{2} (\Gamma_{ij} + H_{ij}) (\Delta\theta^i -\Delta\theta_{\rm bf}^i) (\Delta\theta^j -\Delta\theta_{\rm bf}^j)
 %(\Gamma+H)^{-1}_{ik} (N_k+P_k)) (\Delta\theta^j -(\Gamma+H)^{-1}_{jl} (N_l+P_l)) \right.
 \right.\nonumber \\
 &\hspace{1cm} \left.\left.+ \frac{1}{2} (N_i+P_i)(\Gamma+H)^{-1}_{ij}(N_j+P_j) \right\}\right] \nonumber\\
 &=(2\pi)^{N/2}\frac{p(\vec\theta_0|\vec\lambda) \exp[-(\mathbf{n}|\mathbf{n})/2]}{P_{\rm det}(\vec\lambda) \sqrt{{\rm det}(\Gamma+H)}} \nonumber \\
 &\hspace{0.5cm} \times \exp\left[\frac{1}{2} (N_i+P_i)(\Gamma+H)^{-1}_{ij}(N_j+P_j)\right], \label{eq:poplikeapprox}
 \end{align}
 where we have written
 $$
 \Delta\theta_{\rm bf}^i=(\Gamma+H)^{-1}_{ik} (N_k+P_k).
$$
This is the point at which the likelihood is maximized and hence is the ``best-fit'' point in parameter space. We can now evaluate the population Fisher matrix using the expression
\begin{align}
-(\Gamma_\lambda)_{ij} &= \int \left(\frac{\partial^2 \ln p(\mathbf{d} | \vec\lambda)}{\partial\lambda^i \partial\lambda^j}\right)_{\vec\lambda_t} p(\mathbf{d}|\vec\lambda_t) {\rm d}\mathbf{d} \nonumber \\
&= \int \int \left(\frac{\partial^2 \ln p(\mathbf{d} | \vec\lambda)}{\partial\lambda^i \partial\lambda^j}\right)_{\vec\lambda_t} p(\vec{\theta}_0|\vec\lambda_t) p(\mathbf{n}| \vec\theta_0) {\rm d}\mathbf{n} {\rm d}\vec\theta_0.
\end{align}
In the above the integral over the noise distribution is conditioned on $\vec\theta_0$ because of selection effects. This integral is over all noise realisations that ensure $\mathbf{d} = \mathbf{h}(\vec\theta_0) + \mathbf{n}$ is above the detection threshold. Substituting  Eq.~\eqref{eq:poplikeapprox} into the above we obtain a sequence of terms. To simplify these we carry out the integral over the noise, $\mathbf{n}$. The only terms in Eq.~\eqref{eq:poplikeapprox} that depend on $\mathbf{n}$ are $N_i$ and the prefactor $\exp[-(\mathbf{n}|\mathbf{n})/2]$. The latter enters $\ln \tilde{p}$ additively and has no dependence on the population parameters, so it does not contribute to the final result. The former term also has no explicit dependence on the population parameters, but it appears multiplied by terms that do. There are thus three distinct types of term that appear in the argument of the integral - terms that have no explicit dependence on $\mathbf{n}$, terms that are linear in $N_i$ and terms that are quadratic in $N_i$. We define these integrals as follows
\begin{align}
P_{\rm det} (\vec\theta_0) &= \int p(\mathbf{n} | \vec\theta_0) {\rm d}\mathbf{n} \nonumber \\
\label{eq:Dijk}
D_{i} &\equiv \int \left( \mathbf{n} \bigg| \frac{\partial \mathbf{h}}{\partial\theta^i} \right) p(\mathbf{n}) {\rm d}\mathbf{n} = \frac{\partial P_{\rm det}(\vec\theta_0)}{\partial\theta^i}\,,\nonumber\\ 
D_{ij} &\equiv \int \left( \mathbf{n} \bigg| \frac{\partial \mathbf{h}}{\partial\theta^i} \right) \left( \mathbf{n} \bigg| \frac{\partial \mathbf{h}}{\partial\theta^j} \right) p(\mathbf{n}) {\rm d}\mathbf{n}.
\end{align}
Using these expressions to carry out the integrals over $\mathbf{n}$, we obtain the final result
\begin{align}\label{eq:gammalambda}
(\Gamma_\lambda)_{ij}
&\equiv (\Gamma_\text{I})_{ij}+(\Gamma_\text{II})_{ij}+(\Gamma_\text{III})_{ij}+(\Gamma_\text{IV})_{ij}+(\Gamma_\text{V})_{ij},
\end{align}
with
\begin{align}
(\Gamma_\text{I})_{ij}&=-\int \frac{\partial^2 \ln (p(\vec\theta_0 | \vec\lambda)/P_{\rm det}(\vec\lambda))}{\partial\lambda^i \partial\lambda^j} \, \frac{P_{\rm det}(\vec\theta_0)}{P_{\rm det}(\vec\lambda)} p(\vec\theta_0 | \vec\lambda) {\rm d} \vec\theta_0, \nonumber \\
(\Gamma_\text{II})_{ij}&= \frac{1}{2} \int \frac{\partial^2 \ln {\rm det}(\Gamma+H)}{\partial\lambda^i \partial\lambda^j} \, \frac{P_{\rm det}(\vec\theta_0)}{P_{\rm det}(\vec\lambda)} p(\vec\theta_0 | \vec\lambda) {\rm d} \vec\theta_0,\nonumber \\
(\Gamma_\text{III})_{ij}&= -\frac{1}{2} \int \frac{\partial^2}{\partial\lambda^i \partial\lambda^j}\left[(\Gamma+H)^{-1}_{kl}\right] D_{kl} \, \frac{p(\vec\theta_0 | \vec\lambda)}{P_{\rm det}(\vec\lambda)}  {\rm d} \vec\theta_0, \nonumber \\
(\Gamma_\text{IV})_{ij}&=  -\int \frac{\partial^2}{\partial\lambda^i \partial\lambda^j} \left[ P_k(\Gamma+H)^{-1}_{kl}\right]D_{l} \, \frac{p(\vec\theta_0 | \vec\lambda)}{P_{\rm det}(\vec\lambda)}  {\rm d} \vec\theta_0, \nonumber \\
(\Gamma_\text{V})_{ij}&= -\frac{1}{2} \int \frac{\partial^2}{\partial\lambda^i \partial\lambda^j} \left[ P_k (\Gamma+H)^{-1}_{kl} P_l \right] \, \frac{P_{\rm det}(\vec\theta_0)}{P_{\rm det}(\vec\lambda)} p(\vec\theta_0 | \vec\lambda) {\rm d} \vec\theta_0\nonumber.
\end{align}
This is an approximate expression for the population Fisher matrix which can be used to estimate the precision with which observations will be able to determine the population parameters. {In deriving the above expressions we have made use of the standard form of the likelihood for the gravitational wave detection problem, which permits some simplifications. In Appendix~\ref{app:GenLike} we describe how the result is changed for a generic likelihood, $p(\mathbf{d}|\theta_0)$.}

{We note that, when measurement errors for the source parameters are small, only the first of these terms is required. This limit corresponds to $\Gamma \rightarrow \infty$, so that $\Gamma + H \approx \Gamma$ and $(\Gamma + H)^{-1} \rightarrow 0$. In this limit it is clear that $(\Gamma_\text{III})_{ij}$, $(\Gamma_\text{IV})_{ij}$ and $(\Gamma_\text{V})_{ij}$ immediately vanish. The matrix $(\Gamma_\text{II})_{ij}$ also vanishes because $\Gamma$ does not depend on the population parameters $\lambda$. Therefore we expect $(\Gamma_\text{I})_{ij}$ to dominate and provide a good approximation to the population Fisher matrix. This will be true whenever individual measurement errors are small relative to the scale over which the population parameters change the source parameter distribution. The approximation holds in the three examples we describe below, but this will not always be the case.}
%where the SNR tends to infinity, which means the single source Fisher matrix $\Gamma$ also tends to infinity, the terms  Furthermore, given that the single source Fisher matrix does not depend on the population parameters $\lambda$, . Therefore  for the full Fisher matrix in the case of high SNR, as we will show both analytically and numerically in the examples of the following sections.}
%In the next sections we will illustrate the use of this result for a few sample problems.

\subsection{Validity of approximations}
\label{sec:validity}
To derive the population Fisher matrix we have made two approximations. Firstly, we have used expression~(\ref{eq:asymptotic_cov}) to define the population Fisher matrix. Corrections to this expression are derived in Appendix~\ref{app:popFMcorrect} and are shown to scale with inverse powers of the number of observed events, $n$. This assumption will therefore always be valid once we have made sufficiently many observations, and this is the limit in which we want to use this result. The second approximation was to use the linear signal approximation to represent the posteriors for individual events in Eq.~(\ref{eq:LSA_individ}) and Eq.~(\ref{eq:lnp_P_H}). This approximation will not necessarily be valid in all circumstances, or across the whole of parameter space. In~\citep{Vallisneri:2007ev} a criterion is provided for the validity of the individual event Fisher matrix at $\vec\theta_0$
\begin{align}
    &\frac{1}{2} \left( \Delta\mathbf{h}_{\rm LSA} (\vec{\Delta\theta}) | \Delta\mathbf{h}_{\rm LSA} (\vec{\Delta\theta}) \right) \ll 1 \nonumber \\
    & \hspace{3cm} \forall \vec{\Delta\theta}_{1\sigma} : \Gamma_{ij} \Delta\theta^i_{1\sigma} \Delta\theta^j_{1\sigma} = 1 \nonumber \\
    \mbox{where }& \Delta\mathbf{h}_{\rm LSA} (\vec{\Delta\theta}) = \Delta\theta^i_{1\sigma} \frac{\partial \mathbf{h}}{\partial \theta^i} - \left[ \mathbf{h}(\vec\theta_0+\vec{\Delta\theta}_{1\sigma}) - h(\vec\theta_0)\right] .
\end{align}
If this criterion holds throughout the parameter space of observed events, then the approximations used to derive the population Fisher matrix will definitely be valid. However, this condition is more stringent than is strictly required since the population Fisher matrix is determined by derivatives with respect to the population parameters of the average of the individual event Fisher matrix over the parameter space.

An alternative criterion can be obtained by identifying the next higher order terms in Eq.~(\ref{eq:LSA_individ}) and Eq.~(\ref{eq:lnp_P_H}). These contribute a multiplicative correction to the integral (\ref{eq:poplikeapprox}) of the form
\begin{equation*}
    \exp\left[ \left( \Delta_{ijk}+T_{ijk} \right) \Delta\theta^i \Delta\theta^j \Delta\theta^k + N_{ij} \Delta\theta^i \Delta\theta^j\right],
\end{equation*}
where
\begin{align}
    T_{ijk} &= \frac{1}{6}\frac{\partial^3 \ln p(\vec\theta|\vec\lambda)}{\partial\theta^i \partial\theta^j \partial\theta^k} \qquad
    \Delta_{ijk} =\frac{1}{2} \left(\frac{\partial^2 \mathbf{h}}{\partial\theta^i\partial\theta^j} \bigg| \frac{\partial \mathbf{h}}{\partial\theta^k} \right) \nonumber \\
    N_{ij} &= \left(\frac{\partial^2 \mathbf{h}}{\partial\theta^i\partial\theta^j} \bigg| \mathbf{n} \right).
\end{align}
Approximating the exponential as $\exp(x) \approx 1 + x$, these terms contribute additively to the integral over $\Delta\theta$ an amount $\delta I$. The contribution to $\ln p(\mathbf{d}|\vec\lambda)$ is then an additive $\ln(1 + \delta I/I_0) \approx \delta I/I_0$, where $I_0$ is the value of the leading order integral. We deduce that the next order correction to the population Fisher matrix is
\begin{align}
    \Gamma_{\text{VI}} &= \int \int \left(\frac{\partial^2 (\delta I/I_0)}{\partial\lambda^i \partial\lambda^j}\right)_{\vec\lambda_t} p(\vec{\theta}_0|\vec\lambda_t) p(\mathbf{n}| \vec\theta_0) {\rm d}\mathbf{n} {\rm d}\vec\theta_0, \nonumber \\
    \frac{\delta I}{I_0} &= \left( \Delta_{ijk}+T_{ijk} \right) \left(3 \Delta\theta_{\rm bf}^i (\Gamma+H)^{-1}_{jk} + \Delta\theta_{\rm bf}^i \Delta\theta_{\rm bf}^j \Delta\theta_{\rm bf}^k\right) \nonumber \\
    &\hspace{1cm} + N_{ij}\left( \Delta\theta_{\rm bf}^i \Delta\theta_{\rm bf}^j + (\Gamma+H)^{-1}_{ij}\right). \label{eq:gammVI}
\end{align}
This expression can be simplified further, but as we will not use it elsewhere in this paper we will leave it in this form, but we will make a few observations
\begin{itemize}
    \item This expression can be used to assess the validity of the approximations used to build the population Fisher matrix. If the predicted errors computed including this correction are similar to those computed without then we can trust the population Fisher matrix.
    \item The correction depends on derivatives with respect to the population parameters. If higher order corrections are only significant for parameters that are weakly coupled to those described by the population model, then this correction is still likely to be small, and the predictions of the population Fisher matrix are likely to be trustworthy.
    %If the individual Fisher matrix is only a poor approximation for parameters that , then this correction will be less important and the approximations should be valid.
    \item In the examples discussed later, the dominant contribution to the Fisher matrix comes from $\Gamma_\text{I}$, which is independent of the individual event uncertainties. Thus, even if $\Gamma_\text{VI}$ is of comparable size to $\Gamma_\text{V}$, it might still be negligible relative to $\Gamma_\text{I}$. In that case, $\Gamma_\text{I}$ can continue to be used to estimate the population parameter uncertainties.
    %\item The individual event Fisher matrix and higher corrections appear in combination with the derivatives of the prior, e.g., as $\Gamma+H$. This has a regularizing effect in that the higher order corrections are order unity for low SNR, rather than scaling like inverse powers of SNR.
    \item In general, the Fisher Matrix approximation will be better for events of higher SNR, and so the population Fisher matrix will tend to be a better approximation if we use a higher threshold for including events in the analysis. By adjusting the detection threshold to be high enough that the individual events are well characterised by the Fisher matrix, we will be able to obtain a reliable estimate from the population Fisher matrix for any analysis. This will provide a conservative estimate to the precision that could be achieved using all events.
    %By doing so, we can always use the population Fisher matrix to obtain a conservative estimate of the achievable precision. The precision achieved using all events, which would be done in practice, will be at least as good as estimated using the population Fisher matrix with a high enough detection threshold to ensure the validity of the approximations.
\end{itemize}
In summary, when using the population Fisher matrix to scope out the potential of future GW observations, it is important to monitor the validity of the approximations by using Eq.~(\ref{eq:gammVI}. In some contexts the individual event Fisher matrix will be valid throughout the parameter space of observable events and so the full population Fisher matrix can be used directly. For example, in the context of extreme-mass-ratio inspirals, on which the GW-like example in section~\ref{sec:GWlike_simp} is based, it is expected that a signal-to-noise ratio of at least $20$ will be required for the confident detection of an event in the data~\citep{Babak:2017tow}, and so the Fisher matrix is likely to be a good approximation for all observed events. In other contexts the individual event Fisher matrix might be a poor approximation for some parameters, but if it is valid for the parameters for which the population model has been written down, and these are weakly correlated with the other parameters, then the population Fisher matrix is still likely to provide a good estimate of measurement precision. There will be situations in which the approximation will not be valid, but even there the population Fisher matrix might be accurate if it is dominated by the measurement-error independent part, $\Gamma_\text{I}$. In any scenario, it can be used to obtain a conservative estimate of accuracy by raising the detection threshold sufficiently. IT will thus always provide a valuable tool for quickly scoping out the potential of future observations without the need for expensive simulations.

%higher order terms do not diverge for low SNR - everything O(1)

%summary: check validity. if ok then can use. if not, check if Gamma_I ~ Gamma. If so then can continue to use Gamma_I to assess quality of pop inference

%%%%%%%%%%%%%%%%%%%%%%%%%%%%%%%%%%%%%%%%%%%%%%%%%%
\section{Illustration I: a gaussian-gaussian model} 
%%%%%%%%%%%%%%%%%%%%%%%%%%%%%%%%%%%%%%%%%%%%%%%%%%
\label{sec:gaussian-gaussian}
We will now consider several examples, which will demonstrate that the population Fisher matrix works and show how to compute it in practice. The first application of Eq. \eqref{eq:gammalambda} we will consider is to a ``Gaussian-Gaussian'' model in which both observations and noise are normally distributed. We simplify the setting by assuming a waveform dependent on a single parameter $\theta$. The distribution of the parameter is
\begin{align}\label{pthetalambda}
p(\theta|\vec\lambda)= \mathcal{N}(\mu,\Sigma^2) = \frac{1}{\sqrt{2\pi \Sigma^2}}
\exp\left[-\frac{(\theta-\mu)^2}{2\Sigma^2}\right],
\end{align}
with population parameters (henceforth, hyperparameters) $\vec\lambda =\{\mu,\Sigma^2\}$.
Noise is also a Gaussian with zero mean and variance $\sigma$.
Since the data stream is a sum of Gaussians, it is modelled by $\mathcal{N}(\mu,\sigma^2+\Sigma^2)$, with mean and variance given by the sums of individual means and variances,
\begin{align}\label{eq:pdlambda}
p(\data|\vec\lambda)= \frac{1}{\sqrt{2\pi (\sigma^2+\Sigma^2)}}
\exp\left[-\frac{(\data-\mu)^2}{2(\sigma^2+\Sigma^2)}\right].
\end{align}
Given the implicit simple choice for the signals, the Fisher matrix of source parameters reduces to
\begin{equation}\label{eq:gamma}
\Gamma_\theta = \left(\frac{\partial \boldsymbol h} {\partial \theta}\bigg|\frac{\partial \boldsymbol h} {\partial \theta}\right) = \frac{1}{\sigma^2}\left|\frac{\partial  \boldsymbol h}{\partial \theta}\right|^2=\frac{1}{\sigma^2},
\end{equation}
while from Eq.~\eqref{pthetalambda} we have that $P$ and $H$ in \eqref{eq:lnp_P_H} are
\begin{align}\label{eq:PH}
P = -\frac{(\theta-\mu)}{\Sigma^2} \quad \text{and}\quad
H= \frac{1}{\Sigma^2}\,.
\end{align}
The example reported in this section does not have an immediate analogy in GW astrophysics, but it can be thought of as a more general application of the population Fisher matrix. The advantage of choosing such a simple setting is that the matrix entries can be directly integrated as expectation values over data realizations. In the presence of selection effects, the integrals to be solved are
\begin{align}\label{eq:Gamma_expectation}
  (\Gamma_\lambda)_{ij} &= \mathbb{E} \left(-\frac{\partial^2}{\partial \lambda^i\partial \lambda^j}\left[\ln \left(\frac{p(\boldsymbol d|\theta,\vec\lambda)}{\pdetl}\right)\right]\right) \nonumber\\
  &= - \int_{\dth}^\infty \frac{\partial^2}{\partial \lambda^i\partial \lambda^j}\left[\ln \left(\frac{p(\boldsymbol d|\vec\lambda)}{\pdetl}\right)\right] \frac{p(\boldsymbol d|\vec\lambda)}{\pdetl}d\boldsymbol d.
\end{align}
We only select realizations of the data $\data>\dth$ that are above a certain threshold. The predictions for the various components of the population Fisher matrix are\footnote{For the last integral, it is useful to notice that 
\begin{equation*}
    \int_{\dth}^\infty (\data-\mu)^2 p(\boldsymbol d|\vec\lambda)d\boldsymbol d = \sigmaplussigma \left[1+(\dth-\mu)\frac{p(\dth|\vec\lambda)}{\pdetl}\right],
\end{equation*}
which follows from $(\data-\mu)p(\boldsymbol d|\vec\lambda) = - \sigmaplussigma \partial p(\boldsymbol d|\vec\lambda)/\partial \boldsymbol d$, integrating by parts, and using Eq.\eqref{eq:pdetl}.}
\begin{align}
    (\Gamma_\lambda)_{\mu\mu}&= \frac{\partial^2 \ln \pdetl}{\partial \mu^2}+\frac{1}{\sigmaplussigma},\nonumber\\
    (\Gamma_\lambda)_{\mu\Sigma^2}&=\frac{\partial^2 \ln \pdetl}{\partial \mu\partial \Sigma^2}+ \frac{1}{\sigmaplussigma}\frac{p(\dth|\vec{\lambda})}{\pdetl},\nonumber\\
    (\Gamma_\lambda)_{\Sigma^2\Sigma^2}&=\frac{\partial^2 \ln \pdetl}{(\partial\Sigma^2)^2} + \frac{1}{2\sigmaplussigma^2} \nonumber\\
    &\quad+ \frac{(\dth-\mu)}{\sigmaplussigma^2}\frac{p(\dth|\vec\lambda)}{\pdetl},\label{gamma_mumupred}
\end{align}
where we have used Eqs.\eqref{eq:pdlambda}, \eqref{eq:Gamma_expectation}, the fact that the integral is normalized through
\begin{equation}\label{eq:pdetl}
    \pdetl = \int_{\dth}^\infty p(\boldsymbol d|\vec\lambda)d\boldsymbol d = \frac{1}{2} {\rm erfc} \left(\frac{(\dth-\mu)}{\sqrt{2\sigmaplussigma}}\right),
\end{equation}
and the definition
\begin{equation}\label{pdet_def}
    p(\dth|\vec{\lambda}) \equiv \frac{1}{\sqrt{2\pi \sigmaplussigma}}\exp \left[- \frac{(\dth-\mu)^2}{2\sigmaplussigma}\right].
\end{equation}
Finally, from Eq. \eqref{eq:pdetl} it follows that
\begin{align}
    &\frac{\partial^2 \ln p_\text{det}(\lambda)}{(\partial \mu)^2} =\left(\frac{\dth-\mu}{\sigma^2+\Sigma^2}\right)\frac{p(\dth|\lambda)}{p_\text{det}(\lambda)}-\frac{p(\dth|\lambda)^2}{p_\text{det}(\lambda)^2},\nonumber\\
&\frac{\partial^2 \ln p_\text{det}(\lambda)}{\partial \mu\partial\Sigma^2} =\frac{p(\dth|\lambda)}{p_\text{det}(\lambda)}\left[\frac{(\dth-\mu)^2}{2(\sigma^2+\Sigma^2)^2}-\frac{1}{2(\sigma^2+\Sigma^2)}\right] \nonumber\\
&\qquad \qquad \qquad -\frac{p(\dth|\lambda)^2}{2 p_\text{det}(\lambda)^2}\left(\frac{\dth-\mu}{\sigma^2+\Sigma^2}\right),\nonumber\\
&
\frac{\partial^2 \ln p_\text{det}(\lambda)}{(\partial \Sigma^2)^2}=\frac{p(\dth|\lambda)}{p_\text{det}(\lambda)}\left[\frac{(\dth-\mu)^3}{4(\sigma^2+\Sigma^2)^3}-\frac{3(\dth-\mu)}{4(\sigma^2+\Sigma^2)^2}\right]\nonumber\\
&\qquad \qquad \qquad -\frac{p(\dth|\lambda)^2}{4 p_\text{det}(\lambda)^2}\left(\frac{\dth-\mu}{\sigma^2+\Sigma^2}\right)^2.
\end{align}

\subsection{Solution from the population Fisher matrix} 

% The predictions from the expectation value integrals can be obtained analytically evaluating the working definition of population Fisher matrix derived in Eq. \eqref{eq:gammalambda}. 
% While in this simple setting the procedure is more complicated than directly evaluating the expectation values, we consider it an important consistency check for the formalism.
While in this simple setting the direct evaluation of the Fisher Matrix as expectation value is much simpler, we wish to now evaluate it using Eq. \eqref{eq:gammalambda} as an important sanity check of that general formula. For $(\Gamma_\lambda)_{\mu\mu}$, we notice that the $(\Gamma_\text{II})_{\mu\mu}$, $(\Gamma_\text{III})_{\mu\mu}$ and $(\Gamma_\text{IV})_{\mu\mu}$ vanish. The second and third terms vanish because $\Gamma+H$ does not depend on $\mu$, while the fourth term vanishes because $P$ is only linear in $\mu$ and two derivatives with respect to it are needed for $(\Gamma_\lambda)_{\mu\mu}$. The only terms contributing are therefore $(\Gamma_\text{I})_{\mu\mu}$ and $(\Gamma_\text{V})_{\mu\mu}$, which leads to
\begin{align}
    (\Gamma_\lambda)_{\mu\mu} = & - \int \frac{\partial^2 \ln (p(\theta | \vec\lambda)/\pdetl}{\partial\mu^2} \, \frac{P_{\rm det}(\theta)}{\pdetl} p(\theta | \vec\lambda) d \theta \nonumber\\
    &- \frac{1}{2} \int \frac{\partial^2}{\partial\mu^2} \left[ P^2 (\Gamma+H)^{-1}\right] \, \frac{P_{\rm det}(\theta)}{\pdetl} p(\theta | \vec\lambda) d \theta \nonumber\\
    = &   \frac{\partial^2 \ln \pdetl}{\partial\mu^2} + \frac{1}{\sigmaplussigma},
\end{align}
where we have used $P$, $\Gamma$ and $H$ given in Eqs. \eqref{pthetalambda}, \eqref{eq:gamma} and \eqref{eq:PH}, as well as the normalisation
\begin{equation}\label{eq:pdetnormalization}
    \int \frac{P_{\rm det}(\theta)}{\pdetl} p(\theta | \vec\lambda) d \theta =1.
\end{equation}
The result matches the prediction of Eq.\eqref{gamma_mumupred} as expected. 

In the case of $(\Gamma_\lambda)_{\mu\Sigma^2}$, the second and third terms, $(\Gamma_\text{II})_{\mu\Sigma^2}$ and $(\Gamma_\text{III})_{\mu\Sigma^2}$, vanish for the same reason as above, but now the fourth term $(\Gamma_\text{IV})_{\mu\Sigma^2}$ does contribute. Overall we have that
\begin{align}
    (\Gamma_\lambda&)_{\mu\Sigma^2} =  - \int \frac{\partial^2 \ln (p(\theta | \vec\lambda)/\pdetl}{\partial\mu\partial \Sigma^2} \, \frac{P_{\rm det}(\theta)}{\pdetl} p(\theta | \vec\lambda) d \theta \nonumber\\
    & -\int \frac{\partial^2}{\partial\mu\partial \Sigma^2} \left[ P(\Gamma+H)^{-1}\right]D_{l} \, \frac{P_{\rm det}(\theta)}{\pdetl} p(\theta | \vec\lambda) d \theta\nonumber\\
    &- \frac{1}{2} \int \frac{\partial^2}{\partial\mu\partial \Sigma^2} \left[ P^2 (\Gamma+H)^{-1}\right] \, \frac{P_{\rm det}(\theta)}{\pdetl} p(\theta | \vec\lambda) d \theta \nonumber\\
    = & \frac{\partial^2 \ln \pdetl}{\partial\mu\partial \Sigma^2} + \int \frac{(\theta-\mu)}{\Sigma^4}\, \frac{P_{\rm det}(\theta)}{\pdetl} p(\theta | \vec\lambda) d \theta \nonumber\\
    &+\int  \left[ \frac{\sigma^2}{\sigmaplussigma^2}\right]\frac{\partial P_{\rm det}(\theta)}{\partial\theta} \, \frac{p(\theta | \vec\lambda)}{\pdetl} d \theta\nonumber\\
    & - \frac{\sigma^2(\sigma^2+2\Sigma^2)}{\Sigma^4\sigmaplussigma^2}\frac{1}{\pdetl}\int (\theta-\mu)P_{\rm det}(\theta) p(\theta | \vec\lambda) d \theta, \nonumber
\end{align}
where we have used the definition of $D_{i}$ in Eq. \eqref{eq:Dijk} and \eqref{eq:pdetnormalization}. We also notice that, from the definition
\begin{align}\label{eq:pdet_theta}
    P_{\rm det}(\theta) &= \int_{\dth}^\infty p(\data|\theta) d\data\\
    =&\int_{\dth}^\infty \frac{1}{\sqrt{2\pi\sigma^2}}\exp\left[- \frac{(\data-\theta)^2}{2\sigma^2}\right] d\data = \frac{1}{2}\text{erfc}\left(\frac{\dth-\theta}{\sqrt{2\sigma^2}}\right),\nonumber
\end{align}
it follows that
\begin{align}\label{eq:pdettheta}
    \frac{\partial P_{\rm det}(\theta)}{\partial\theta} &= \int_{\dth}^\infty \frac{(\data-\theta)}{\sigma^2}p(\data|\theta) d\data \nonumber\\
    =& \frac{1}{\sqrt{2\pi \sigma^2}} \exp\left[- \frac{(\dth-\theta)^2}{2\sigma^2}\right] \equiv p(\dth|\theta),
\end{align}
which can be rearranged to give
\begin{align}
    \label{eq:constraint}
    \int (\theta -\mu)&\frac{\partial P_{\rm det}(\theta)}{\partial \theta} p(\theta | \vec\lambda) d \theta \\
& = \Sigma^2 \frac{\partial}{\partial \mu}\int p(\dth|\theta) p(\theta|\vec \lambda)d\theta \nonumber
\\
&= \Sigma^2 \frac{\partial p(\dth|\vec \lambda)}{\partial \mu} = \frac{\Sigma^2}{\sigmaplussigma}(\dth-\mu)p(\dth|\vec \lambda).\nonumber
\end{align}
From Eq. \eqref{eq:pdetl} it also follows that $\partial\pdetl/\partial \mu = p(\dth|\vec\lambda)$. Using this constraint, and the fact that
\begin{align}
\frac{\partial p(\dth|\vec\lambda)}{\partial \mu} &= \int \frac{\partial}{\partial\mu} \big[P_{\rm det}(\theta) p(\theta|\vec\lambda)\big]d\theta \nonumber\\
&= \frac{1}{\Sigma^2}\int (\theta -\mu)P_{\rm det}(\theta) p(\theta | \vec\lambda) d \theta,
\end{align}
we immediately get
\begin{equation}\label{eq:theta_meno_mu}
    \int (\theta -\mu)P_{\rm det}(\theta) p(\theta | \vec\lambda) d \theta = \Sigma^2 p(\dth|\vec\lambda),
\end{equation}
and, together with \eqref{eq:pdetnormalization} and \eqref{eq:constraint}, that 
\begin{equation}
(\Gamma_\lambda)_{\mu\Sigma^2} 
%=\frac{\partial^2 \ln \pdetl}{\partial \mu\partial \Sigma^2}+ \int_{\dth}^\infty \frac{(\data-\mu)}{\sigmaplussigma} \frac{p(\boldsymbol d|\vec\lambda)}{\pdetl}d\boldsymbol d 
= \frac{\partial^2 \ln \pdetl}{\partial \mu\partial \Sigma^2}+ \frac{1}{\sigmaplussigma}\frac{p(\dth|\vec{\lambda})}{\pdetl}.
\end{equation}
This agrees with Eq. \eqref{gamma_mumupred} as expected.

Finally, we consider the case of $(\Gamma_\lambda)_{\Sigma^2\Sigma^2}$, in which no term in \eqref{eq:gammalambda} vanishes. The first term can be rearranged to give\footnote{The integral over $\theta$ can be evaluated by parts to obtain
\begin{align}\label{eq:theta_meno_mu_squared}
    \int (\theta-\mu)^2 \,& \frac{P_{\rm det}(\theta)}{\pdetl} p(\theta | \vec\lambda) d \theta = \Sigma^2 + \frac{(\dth-\mu)\Sigma^4}{\sigmaplussigma} \frac{p(\dth|\vec\lambda)}{\pdetl}, 
\end{align}
using \eqref{eq:pdetl} and \eqref{eq:theta_meno_mu} with vanishing boundary terms. }
\begin{align}\label{eq:first_gammass}
    (\Gamma_\text{I})_{\Sigma^2\Sigma^2}&= \frac{\partial^2 \ln \pdetl}{(\partial\Sigma^2)^2} -\frac{1}{2\Sigma^4}\\
    &+\frac{1}{\Sigma^6}\int (\theta-\mu)^2 \, \frac{P_{\rm det}(\theta)}{\pdetl} p(\theta | \vec\lambda) d \theta\nonumber\\
    &=\frac{\partial^2 \ln \pdetl}{(\partial\Sigma^2)^2} +\frac{1}{2\Sigma^4}+ \frac{(\dth-\mu)}{\Sigma^2\sigmaplussigma} \frac{p(\dth|\vec\lambda)}{\pdetl}.\nonumber
\end{align}
Using Eq. \eqref{eq:pdetnormalization}, the second term $(\Gamma_\text{II})_{\Sigma^2\Sigma^2}$ is easily found to be
\begin{equation}\label{eq:second_gammass}
    (\Gamma_\text{II})_{\Sigma^2\Sigma^2} = \frac{\sigma^2}{2\Sigma^4} \frac{(\sigma^2+2\Sigma^2)}{\sigmaplussigma^2}.
\end{equation}
The third term $(\Gamma_\text{III})_{\Sigma^2\Sigma^2}$ contains $D_{ij}$ from Eq.\eqref{eq:Dijk}, which can be rearranged to give
\begin{align}
    D&_{\theta\theta}= \frac{1}{\sqrt{2\pi}\sigma^3}\int_{\dth}^\infty (\data-\theta)^2 \exp\left[-\frac{(\data-\theta)^2}{2\sigma^2}\right]d\data\nonumber\\
    &= \frac{(\dth-\theta)}{\sqrt{2\pi}\sigma^3}\exp\left[-\frac{(\dth-\theta)^2}{2\sigma^2}\right] +\frac{1}{2\sigma^2}\text{erfc}\left(\frac{\dth-\theta}{\sqrt{2\sigma^2}}\right).\nonumber
\end{align}
The complementary error function can be replaced by $P_{\rm det}(\theta)$ using \eqref{eq:pdet_theta}. With this in mind, we have
\begin{align}
    \label{eq:third_gammass}
    (\Gamma_\text{III})_{\Sigma^2\Sigma^2} = \frac{\sigma^4}{\sigmaplussigma^4} (\dth-\mu)\frac{p(\dth|\vec\lambda)}{\pdetl}  +\frac{\sigma^2}{\sigmaplussigma^3}.
\end{align}
The results follows from rearranging Eqs. \eqref{eq:pdettheta} and \eqref{pdet_def} into
\begin{align*}
    &\int (\dth-\theta)\exp\left[-\frac{(\dth-\mu)^2}{2\sigma^2}\right] \, \frac{p(\theta | \vec\lambda)}{\pdetl} d \theta \nonumber\\
    &\qquad\qquad= -\sigma^3 \sqrt{2\pi} \int \frac{\partial p(\dth|\theta)}{\partial \dth}\frac{p(\theta | \vec\lambda)}{\pdetl} d \theta\nonumber\\
    &\qquad\qquad = -\frac{\sigma^3 \sqrt{2\pi}}{\pdetl} \int \frac{\partial}{\partial \dth}\big[p(\dth|\theta)p(\theta | \vec\lambda) \big]d\theta \nonumber\\
    &\qquad\qquad = \frac{\sigma^3 \sqrt{2\pi}}{\sigmaplussigma}(\dth-\mu)\frac{p(\dth|\vec\lambda)}{\pdetl}.
    \nonumber
\end{align*}
The fourth term $(\Gamma_\text{IV})_{\Sigma^2\Sigma^2}$ is found using \eqref{eq:theta_meno_mu}, and reads
\begin{align}\label{eq:fourth_gammass}
    (\Gamma_\text{IV})_{\Sigma^2\Sigma^2}&-  \int \frac{\partial^2}{(\partial\Sigma^2)^2} \left[ P(\Gamma+H)^{-1}\right]D_{\theta} \, \frac{P_{\rm det}(\theta)}{\pdetl} p(\theta | \vec\lambda) d \theta \nonumber\\
    &= \frac{2\sigma^2\Sigma^2}{\sigmaplussigma^4} (\dth-\mu) \frac{p(\dth|\vec\lambda)}{\pdetl}.
\end{align}
Finally, the final term $(\Gamma_\text{V})_{\Sigma^2\Sigma^2}$ is found through \eqref{eq:theta_meno_mu_squared} to be
\begin{align}\label{eq:fifth_gammass}
    (\Gamma_\text{V})_{\Sigma^2\Sigma^2}&=- \frac{1}{2} \int \frac{\partial^2}{(\partial\Sigma^2)^2} \left[ P^2 (\Gamma+H)^{-1}\right]\, \frac{P_{\rm det}(\theta)}{\pdetl} p(\theta | \vec\lambda) d \theta \nonumber\\
    &= -\frac{\sigma^2}{\Sigma^4} \frac{(\sigma^4 + 3\sigma^2 \Sigma^2 + 3\Sigma^4)}{\sigmaplussigma^3}\nonumber\\
    &\qquad\left[1+ \frac{\Sigma^2}{\sigmaplussigma}(\dth-\mu)\frac{p(\dth|\vec\lambda)}{\pdetl}\right].
\end{align}
Adding up \eqref{eq:first_gammass} to \eqref{eq:fifth_gammass}, we find
\begin{align}
    \Gamma_{\Sigma^2\Sigma^2}= \frac{\partial^2 \ln \pdetl}{(\partial\Sigma^2)^2} + \frac{1}{2\sigmaplussigma^2} + \frac{(\dth-\mu)}{\sigmaplussigma^2}\frac{p(\dth|\vec\lambda)}{\pdetl},
\end{align}
which matches \eqref{gamma_mumupred} as expected. This concludes the analytic check. Equation \eqref{eq:gammalambda} can be used to reproduce the predictions \eqref{gamma_mumupred} obtained from the more general definition of the Fisher matrix as an expectation value over data realizations.

\subsection{MCMC analysis}

\begin{figure}
    \includegraphics[width=.5\textwidth]{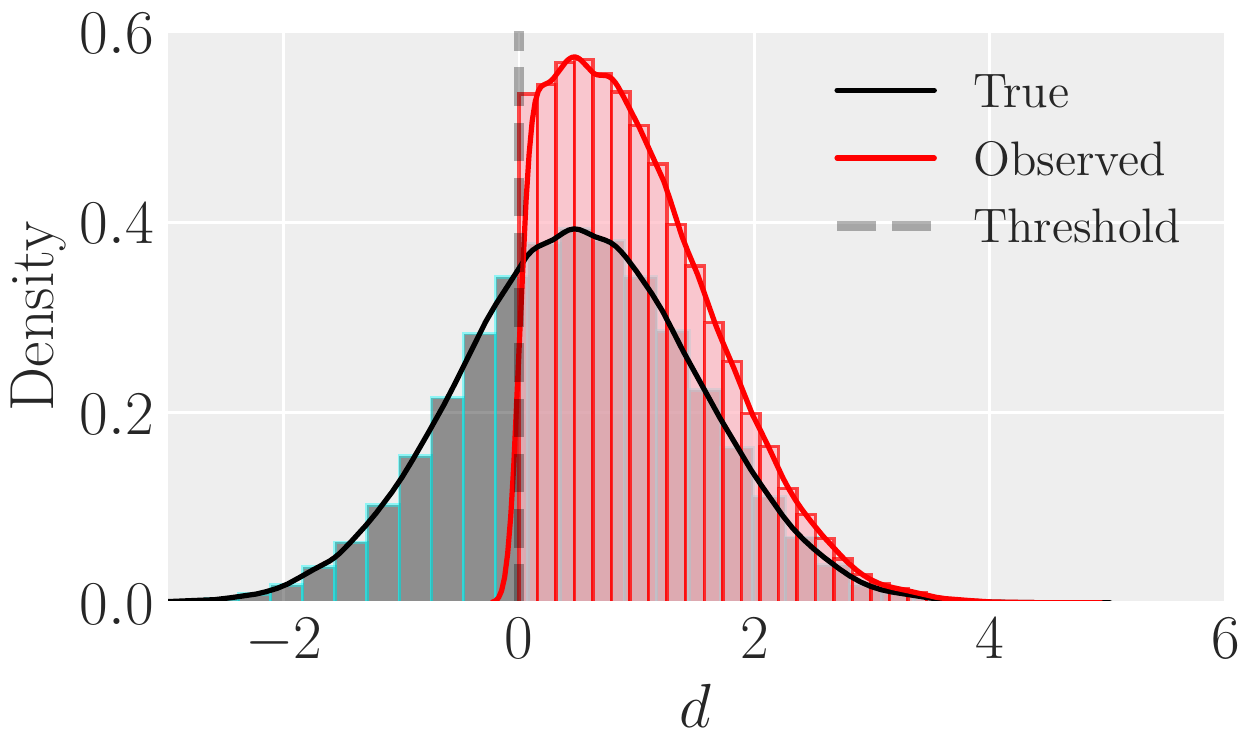}
    \caption{Normally-distributed data with (red) and without (black) selection effects.}\label{fig:gaussian_data}
\end{figure}

\begin{figure}
    \centering
    \includegraphics[width=.49\textwidth]{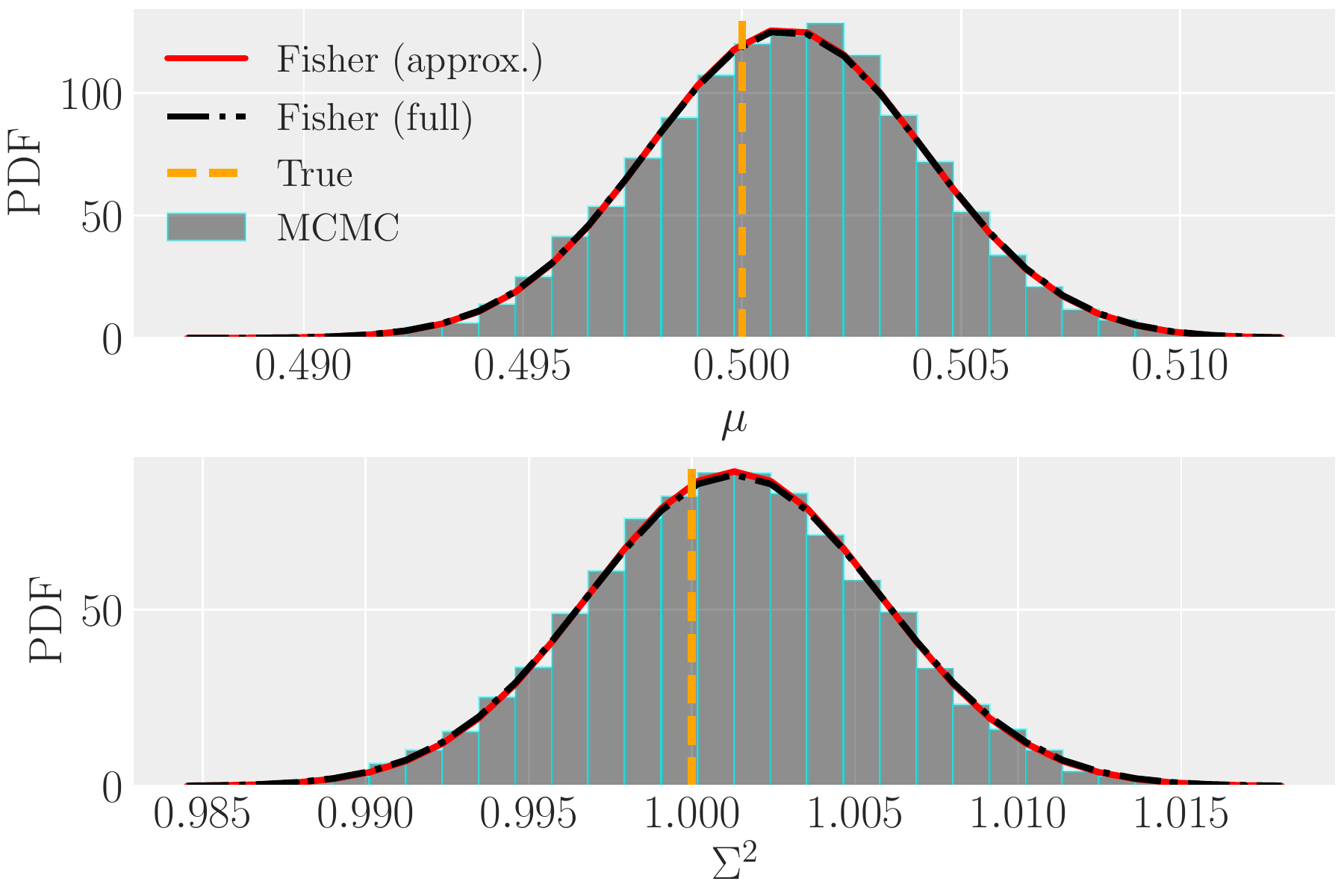}
    \includegraphics[width=.49\textwidth]{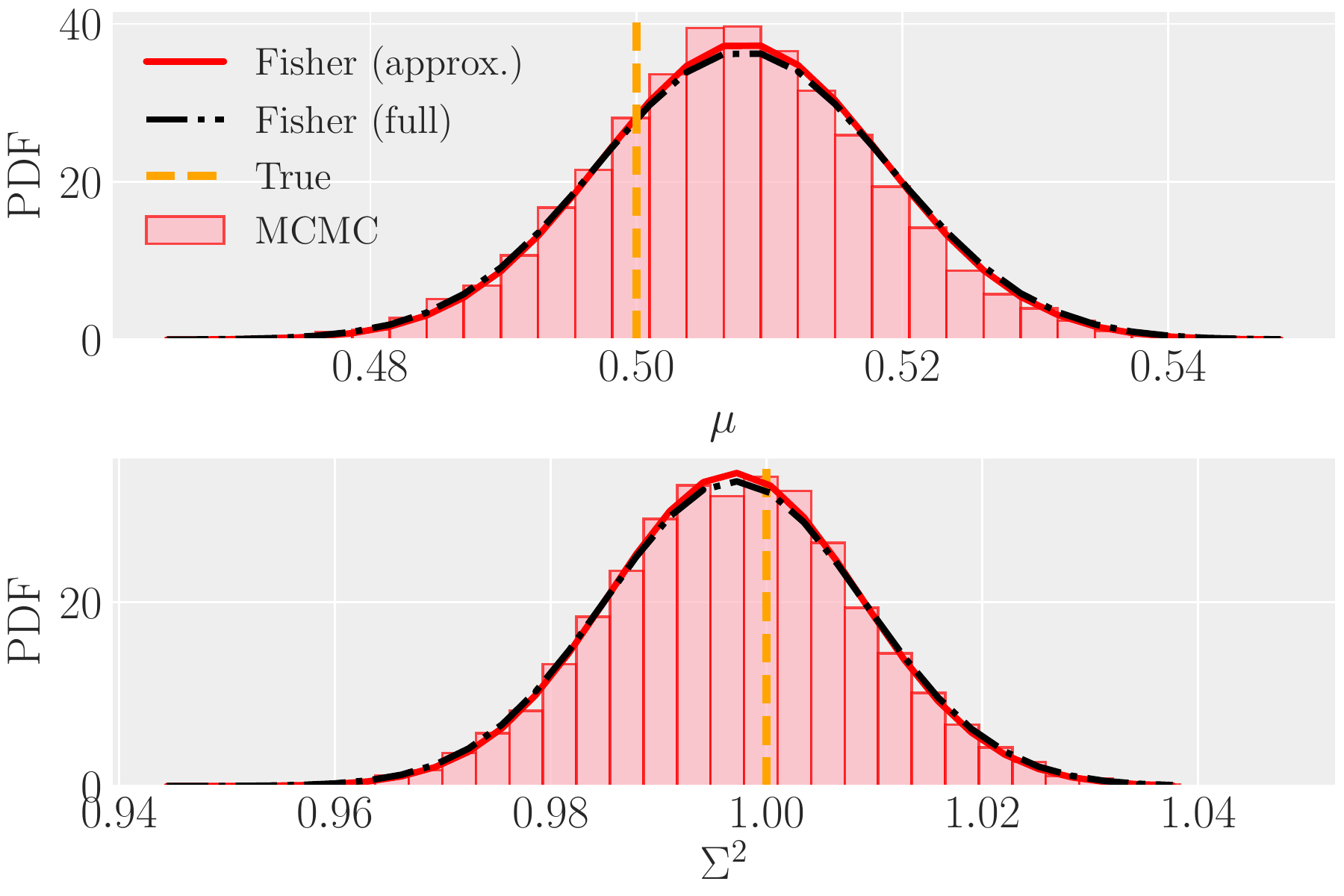}
    \caption{Fisher and MCMC predictions for the toy Gaussian case with (red) and without (black) selection effects.}\label{fig:MCMC_gaussian}
\end{figure}

The Fisher predictions for the Gaussian-Gaussian model can be compared with MCMC simulations as a further check of the formalism. While this example is arguably textbook material, see for instance Sec. (6) in \citep{Vitale:2020aaz}, we report a few details below for completeness. The results of the present section can be fully reproduced with the codes accompanying the present paper.
We simulate synthetic data including $N_\text{tot}=10^5$ observations from the observation model \eqref{eq:pdlambda}, choosing true mean $\mu_\text{tr}=0.5$, true variance $\Sigma^2_\text{tr}=1.0$ and noise variance $\sigma=0.1$. The latter two indicate that each event is taken with a high SNR.
We then apply an arbitrary cutoff, imposing that only positive data are observed. That is, $\boldsymbol d_\text{th}=0$, resulting in around $N_\text{det}\sim 70000$ detected events. 
Figure \ref{fig:gaussian_data} shows the total (black) and detected (red) populations under our specified assumptions.
We perform the MCMC analyses in both cases in which we do and do not have selection effects using \texttt{emcee} \citep{Foreman-Mackey:2012any}. As log-likelihood, we take the sum of individual log-likelihoods,
\begin{align}
    \log p(d|\lambda)=- N\log p_\text{det}(\lambda) + \sum_{i}^{N} \log p(d_i|\lambda),
\end{align}
where $N=N_\text{tot}$ in the case in which we do not include selection effects, and $N=N_\text{det}$ in the case in which we do. We choose flat hyperpriors over a very broad range that includes the true values.
The selection function is defined and integrated as in Eq. \eqref{eq:pdetl}, and is nontrivial only in the latter case. The MCMC posteriors for $\mu$ and $\Sigma^2$, in both the cases considered, can be found in Fig. \ref{fig:MCMC_gaussian}.

We can then compare the predictions from the population Fisher matrix with what we obtain numerically. To this end, we invert the matrix
\begin{equation}\label{eq:Fisher_tot}
    (\Gamma_\lambda)_{ij} = \begin{pmatrix}
\Gamma_{\mu\mu} & \Gamma_{\mu\Sigma^2}\\
\Gamma_{\mu\Sigma^2} & \Gamma_{\Sigma^2\Sigma^2}
\end{pmatrix}
\end{equation}
with entries given in Eq. \eqref{gamma_mumupred} and below.
The errors are normalized by the number of events, 
\begin{equation*}
    \Delta\mu =\sqrt{(\Gamma^{-1}_\lambda)_{\mu\mu}/N}, \quad \Delta\Sigma^2 =\sqrt{(\Gamma^{-1}_\lambda)_{\Sigma^2\Sigma^2}/N}\nonumber.
\end{equation*}
% $\Delta\mu =\sqrt{N(\Gamma^{-1}_\lambda)_{\mu\mu}}$ and $\Delta\Sigma^2 =\sqrt{N(\Gamma^{-1}_\lambda)_{\Sigma^2\Sigma^2}}$. 
In Fig. \ref{fig:MCMC_gaussian}, the Fisher predictions are shown (in black) to reproduce the widths from the MCMC runs. 

The same widths can be well approximated by the inverse of the matrix
\begin{equation}\label{eq:Fisher_I}
    (\Gamma_\lambda)_{ij} = \begin{pmatrix}
(\Gamma_\text{I})_{\mu\mu} & (\Gamma_\text{I})_{\mu\Sigma^2}\\
(\Gamma_\text{I})_{\mu\Sigma^2} & (\Gamma_\text{I})_{\Sigma^2\Sigma^2},
\end{pmatrix}
\end{equation}
in which only the first terms in Eq. \eqref{eq:gammalambda} are retained. The predictions are shown in red in Fig. \ref{fig:MCMC_gaussian}, and they overlap well with the full Fisher matrix predictions. {As discussed earlier, this corresponds to taking the limit of Eq. \eqref{eq:Fisher_tot} in which $\sigma = 0$, i.e., the parameters of the individual events are measured perfectly. The reason that this is a good approximation here is because we have chosen $0.1 = \sigma \ll \Sigma = 1$ for this example. We would expect the other terms contributing to Eq.~\eqref{eq:gammalambda} to become increasingly important as the measurement errors become larger. If we consider for simplicity the case without selection effects we see that the ratio of the uncertainties in the population parameters computed using only $\Gamma_\text{I}$ to that computed using the full Fisher matrix, Eq. \eqref{eq:Fisher_tot}, are}
\begin{align}
\frac{\Delta \mu_\text{I}}{\Delta \mu_\text{full}} &= \frac{1}{1 + \sigma^2/\Sigma^2} \qquad
\frac{\Delta \Sigma^2_\text{I}}{\Delta \Sigma^2_\text{full}} = \frac{1}{1 + \left(\sigma^2/\Sigma^2\right)^2}.
\end{align}
{For $\sigma^2 \ll \Sigma^2$ these ratios are approximately $1$, as expected, but as $\sigma^2/\Sigma^2 \rightarrow 
\infty$ both ratios tend to $0$, implying that $\Gamma_\text{I}$ would significantly over-estimate the precision with which the population parameters can be determined. So, it is not always possible to use $\Gamma_\text{I}$ to estimate the population parameter uncertainties. However, in many applications individual events are constrained to a small region of the much larger parameter space of the population, and so Eq.~\eqref{eq:Fisher_I} will often be a good approximation to the full Fisher matrix. This includes the GW-like illustrations we will consider in the next section.}

%%%%%%%%%%%%%%%%%%%%%%%%%%%%%%%%%%%%%%%%%%%%%%%%%%
\section{Illustration II: an example from gravitational-wave astrophysics} 
%%%%%%%%%%%%%%%%%%%%%%%%%%%%%%%%%%%%%%%%%%%%%%%%%%
\label{sec:EMRI-model}

The spaceborne LISA mission is expected to detect extreme mass ratio inspirals (EMRIs), namely binary systems in which one compact object, typically a stellar remnant, has a mass that is much smaller than the companion, typically a supermassive black hole (SMBH) in the centre of a galaxy \citep{Amaro-Seoane:2007osp, Barack:2009ux, Babak:2017tow}. Inference of the parameters that characterise EMRI systems is expected to provide accurate constraints on the theory of gravity \citep{Gair:2012nm}, as well as an insight into the astrophysical population of and stellar environments surrounding SMBHs \citep{Barausse:2014tra}. LISA might also detect a foreground generated by individually unresolved EMRIs, which would porvide information about the properties of the population of these systems \citep{Gair:2010bx,Gair:2010yu,Sesana:2010wy,Bonetti:2020jku}. The use of LISA observations of EMRIs to provide measurements of the BH mass function in the range probed by LISA has previously been investigated in  \citep{Gair:2010yu}, henceforth ``GTV''. GTV assumed that the mass function was described by a power law $p(\theta|\lambda)\equiv p(M|\lambda)\propto M^{\alpha-1}$, with a true value that is close to flat in the log of the masses, i.e., $\alpha \approx 0$. GTV explored the ability of LISA to constrain the parameters of this mass function, using MCMC techniques to carry out hierarchical analyses on an extensive set of populations of simulated events. They found that with 10(1000) events, the spectral index $\alpha$ could be constrained at a level of precision $\Delta\alpha = 0.3 (0.03)$. %The analysis in GTV was based on an extensive set of MCMC simulations. 
Here, we will use the population Fisher Matrix formalism described above to predict the precision with which a set of EMRI observations might be able to constrain a power-law mass function. We do not expect to get exactly the same answer here, as the two analyses make a few different simplifying assumptions. In GTV the raw data was taken to be counts of events in a binned analysis, provided by point estimates of the parameters. Selection effects were included in the rate of events in each bin, by accounting for the length of time a source with the given parameters would be observable. This ignores the fact that the time remaining to plunge is constrained by the gravitational wave data. In this analysis we again approximate the observation process, assuming that the data can be reduced to a measurement of a single parameter, but we handle selection effects more carefully. GTV's results was computed ignoring measurement uncertainties in the model used in the analysis, although they did demonstrate consistency between results obtained on simulated data with and without measurement uncertainties. Here we will include measurement uncertainties, but we will approximate these as Gaussian. %, with equal variance for all observations. 
We will see that despite these differences in assumptions, the population Fisher matrix is able to predict qualitatively the results observed in GTV without the need for costly computational sampling of many posterior distributions. For a more direct comparison with numerical results we also perform our own MCMC analysis, under identical assumptions to those used to compute the population Fisher matrix, and find very good agreement. %We follow the setup of GTV but treat selection effects consistently with the population Fisher matrix, finding better agreement between the estimates. 
All the results in this section can be reproduced with the codes made available with this publication.

\subsection{Simple scenario: mass measurement only}
\label{sec:GWlike_simp}
\begin{figure*}
\centering
\includegraphics[width=0.8\textwidth]{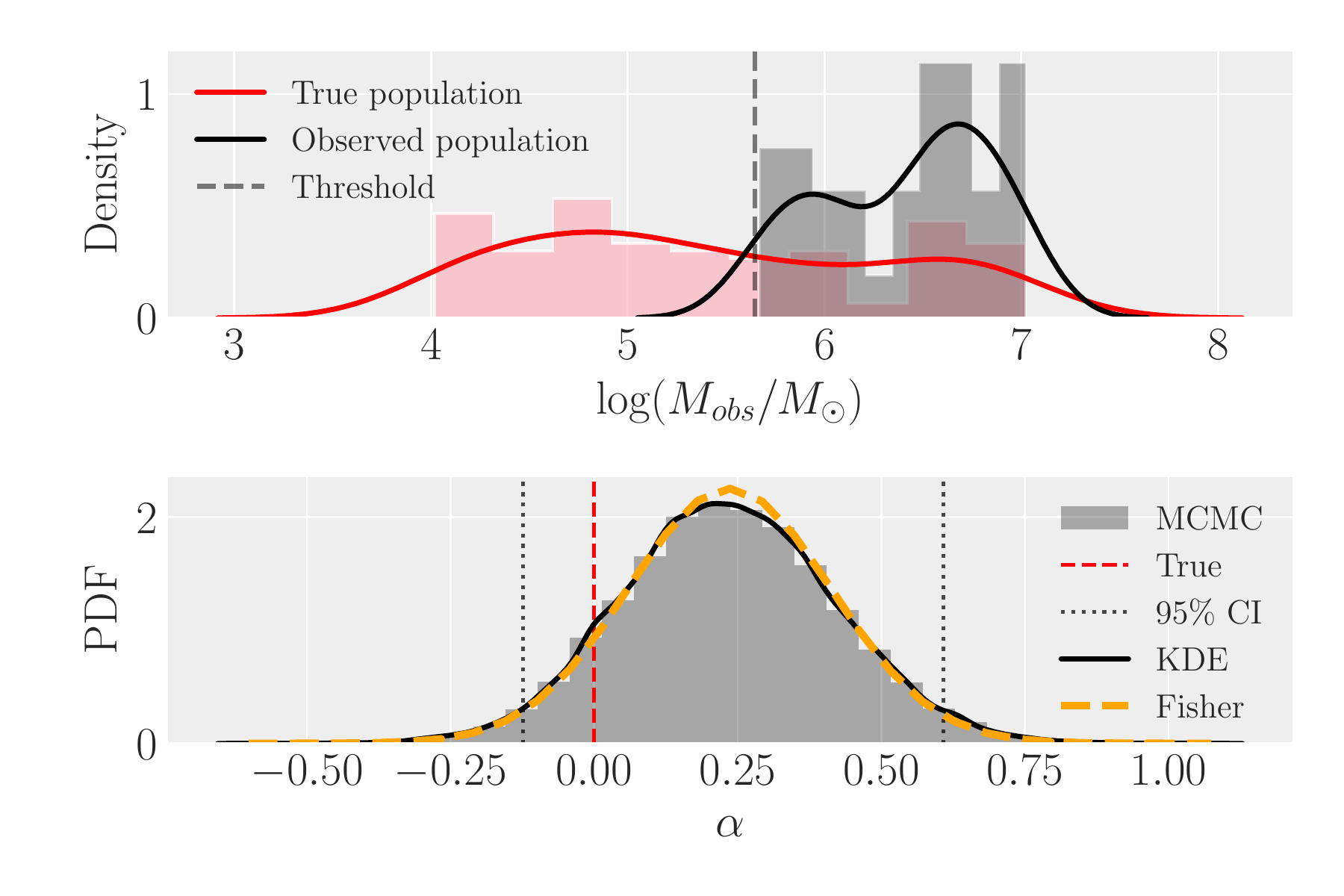}
\caption{\emph{Top panel:} distribution of masses for the underlying true population of Sec. \ref{sec:EMRI-model}  (red) and for the observed population (dark gray). The true population is composed of $N=100$ events drawn from a power-law model that is flat in the logarithm of the masses. The threshold has been arbitrarily set at $\sim 5\times 10^5 M_\odot$, leading to 39 events actually being observed. 
\emph{Bottom panel:} MCMC posterior distribution for the spectral index describing the mass distribution. % in the upper panel. 
The histogram and KDE are compared against the Fisher estimate obtained as described in the text, demonstrating very good agreement between the two.}
\label{fig:GWlike_example}
\end{figure*}

{In the first scenario we consider we will assume that events in the population are characterised by a single parameter, the mass, which we measure with our detector with a Gaussian uncertainty that has a fixed variance, independent of the parameters of the source. We note that this choice of a constant variance is made for convenience, but is not required by the formalism. If errors vary form event to event, these are characterised by a $\vec\theta_0$ dependence of $\Gamma$, which just changes the integrands of the various components of the population Fisher matrix. Due to the integration over $\vec\theta_0$, the population Fisher matrix effectively depends on the average measurement precision over the population. In the final example, described in section~\ref{sec:GWlike_real}, we will consider a case in which the errors vary from event to event.} We draw $N=100$ masses from a power law distribution, 
\begin{equation}\label{eq:power_law_model}
    p(M|\alpha) = \frac{\alpha}{M^\alpha_\text{max}-M^\alpha_\text{min}} M^{\alpha-1},
\end{equation}
with maximal and minimal observable masses $M_\text{min}=10^4 M_\odot$ and $M_\text{max}=10^7 M_\odot$ and a true value for the spectral index that is exactly flat in the log of the masses, $\alpha=0$. The observed data is assumed to be a point estimate of the log of the mass, which is equal to the true value plus a normally-distributed uncertainty that has a variance $\sigma =0.1$. 
% \jg{Are we really using sigma = 0.1 in mass, not log(mass)? This would definitely make the measurement uncertainties negligible! Can you confirm/correct?} \andrea{ops, we are adding uncertainty to the logs of the masses.} 
An arbitrary hard cutoff $\textbf{d}_\text{th}$ corresponding to masses $M\sim 5\times 10^{5} M_\odot$ is imposed, and only events with observed values above this threshold are included in the analysis. This introduces a large selection effect, and leads to only $N_\text{det}=39$ of the original 100 sources being observed. 
%As the selection function is a property of the data observation model, which is kept unchanged from the previous example, we include such selection effects in the Fisher matrix (and likelihood in the MCMC analysis) using YY.
The true (underlying) and observed populations of events are represented in the top panel of Fig.\ref{fig:GWlike_example}.
%JG TO HERE

The Fisher-matrix prediction can again be obtained from Eq. \eqref{eq:gammalambda}, this time with $\theta =M$ and $\lambda =\alpha$. For sufficiently simple models, the integrals are analytically tractable. With the power-law distribution considered here, and in the presence of selection effects, this is already not possible. Because of this, we obtain the Fisher prediction in a semi-analytical fashion by solving the integrals with Monte-Carlo methods, i.e., by generating a sufficiently large set of $N_\text{s}$ samples, $\{M_i\}$, from a distribution $p(M|\alpha)$, we can approximate the integral of an arbitrary function $X(M)$ via
\begin{equation}
    \int X(M) p(M|\alpha)dM\approx \frac{1}{N_\text{s}} \sum_i X(M_i).
\end{equation}
%where $N_\text{s}$ is a sufficiently large number of samples, and $X(M)$ are analytical expressions obtained solving for the arguments of the integrals. 
In this case we can draw samples from the distribution~\eqref{eq:power_law_model} directly using the method of inversion. The various terms entering the arguments of Eq.~\eqref{eq:gammalambda} --- $D_{i}$, $D_{ij}$, $P_i$ and $H_{ij}$ --- can be computed analytically. A \texttt{Mathematica} notebook that solves for the arguments of the Fisher-matrix integrals can be found in the accompanying codes. Following this procedure, we find that the Fisher matrix predicts an error $\Delta \alpha = \sqrt{(N_\text{det}\Gamma_{\alpha})^{-1}}\approx 0.19$. When rescaled to 10 observations by multiplying by $\sqrt{39/10}$, the inferred error is $\Delta \alpha\approx 0.37$, which is in good agreement with what was obtained in GTV, despite the differences in the assumptions used in each case.

\begin{figure*}
\centering
\includegraphics[width=0.8\textwidth]{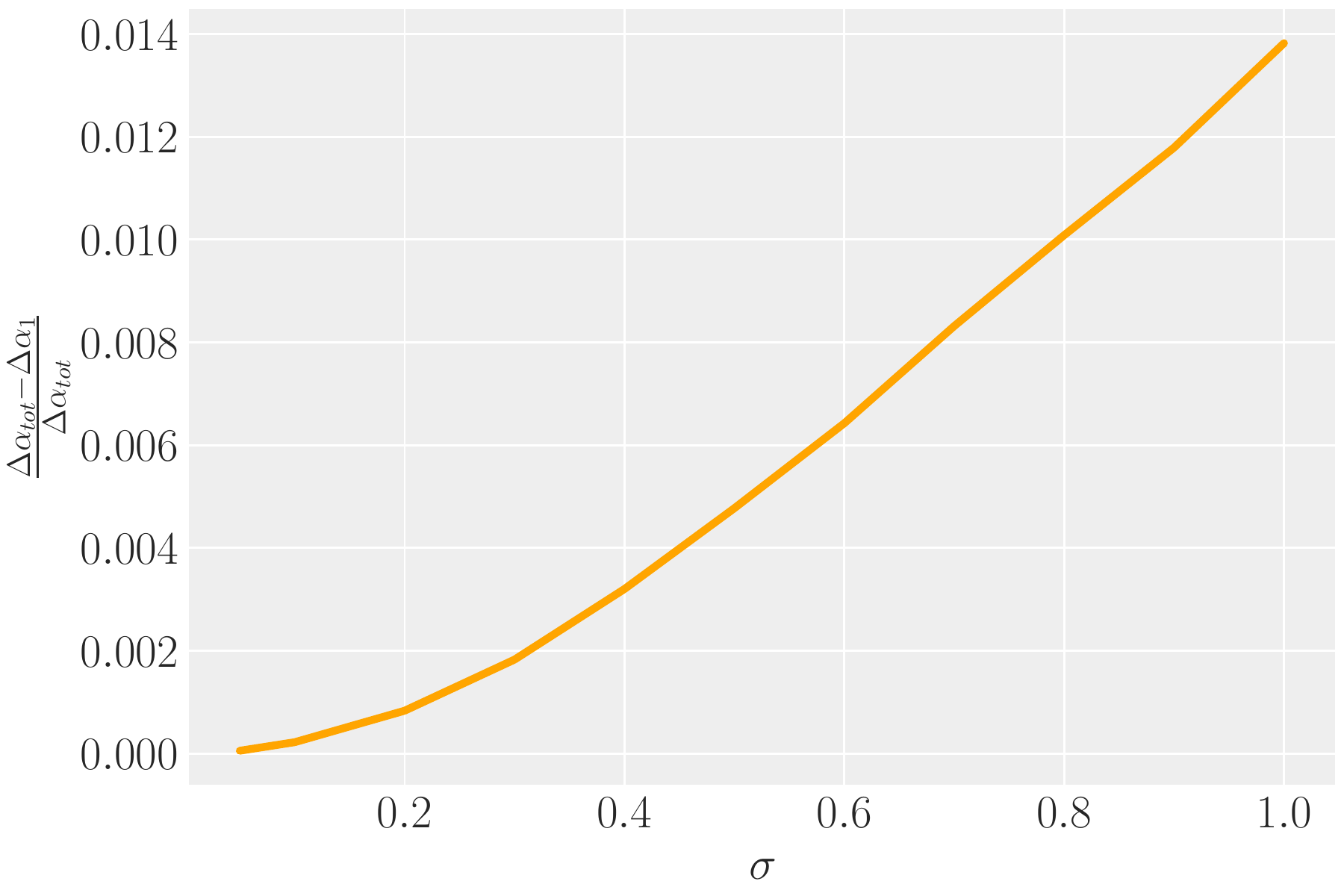}
\caption{Fractional error in the estimate of the precision on the population slope, $\alpha$, obtained from using only the first term in the Fisher matrix, $\Gamma_\text{I}$, instead of the full Fisher matrix, as a function of the assumed uncertainty on the measurement of individual event parameters, $\sigma$.}
\label{fig:ApproxAccVsSigma}
\end{figure*}

As in the Gaussian-Gaussian example, we find that the dominant contribution comes from the first term of the Fisher matrix, $\Gamma_\text{I}$. In this example this term can be directly computed
\begin{equation}\label{eq:Fisher_GWlike_example_analytical}
    \Gamma_\alpha \approx N_\text{tot}\left(\frac{1}{\alpha^2}-\frac{M_\text{max}^\alpha M_\text{min}^\alpha (\ln M_\text{max}-\ln M_\text{min})^2}{(M_\text{max}^\alpha-M_\text{min}^\alpha)^2}\right).
\end{equation}
{As argued above, the dominance of this term is driven by the assumed precision of measurement on the individual events. If the noise in the individual measurements, $\sigma$, is increased, the other terms make a larger contribution, although always a sub-dominant contribution for the range of values we have tried. This is illustrated in Figure~\ref{fig:ApproxAccVsSigma}, which shows how the error in the prediction for the precision on the slope from using only the first term of the Fisher matrix varies with $\sigma$. Even for $\sigma = 1$, the fractional error from this approximation is only $0.014$. This fact suggests that, when individual events are expected to be characterised with a precision better than the typical lengthscale over which the population prior varies, keeping only the first term in the sum will provide a good estimate for the expected precision of population inference, regardless of the population model chosen. This observation and the fact that the first term is typically relatively easy to evaluate, could help to reduce the complexity of using our formalism. One obvious application would be to forecast studies for future detectors, where this formalism for the precision of population inference can nicely complement estimates for the precision of individual event parameter inference, estimated with state-of-the-art Fisher codes \citep{Borhanian:2020ypi,Harms:2022ymm}.} 

%in situations in which individual events are accurately resolved, the first integral is a good proxy for estimates of the expected precision in population inference, 
%which confirms the suggestion made in the example that the first term in the population Fisher matrix is enough to estimate the likelihood. The other integrals only contribute in this case at a precision lower than the digits reported, and their contribution only starts increasing to an appreciable, though still subdominant level when the noise variance is increased. This fact suggests that, in situations in which individual events are accurately resolved, the first integral is a good proxy for estimates of the expected precision in population inference, regardless of the population model chosen. 

We validate the Fisher predictions with an MCMC analysis for the same data set. The MCMC setup is similar to the one used in the first illustrative example. The likelihood is modified, and the selection function is no longer known analytically, but is approximated by a Monte Carlo integral, $p_\text{det}(\alpha)=(1/N_\text{s}) \sum \text{erfc}[(\dth - M_i)/\sqrt{2\sigma^2}]/2$, with $\{M_i\}$ drawn from the power law distribution $p(M|\alpha)$. The posterior, KDE and $2\sigma$ percentile for the estimate of $\alpha$ are shown in the bottom panel of Fig.\eqref{fig:GWlike_example}. These are compared with the Fisher predictions above, and once again show very good agreement. We have also repeated the calculation in the absence of selection effects, $\textbf{d}_\text{th}\rightarrow - \infty$, and find a similar level of agreement between the Fisher predictions and the MCMC analysis. 

\begin{figure}
\centering
\includegraphics[width=\linewidth]{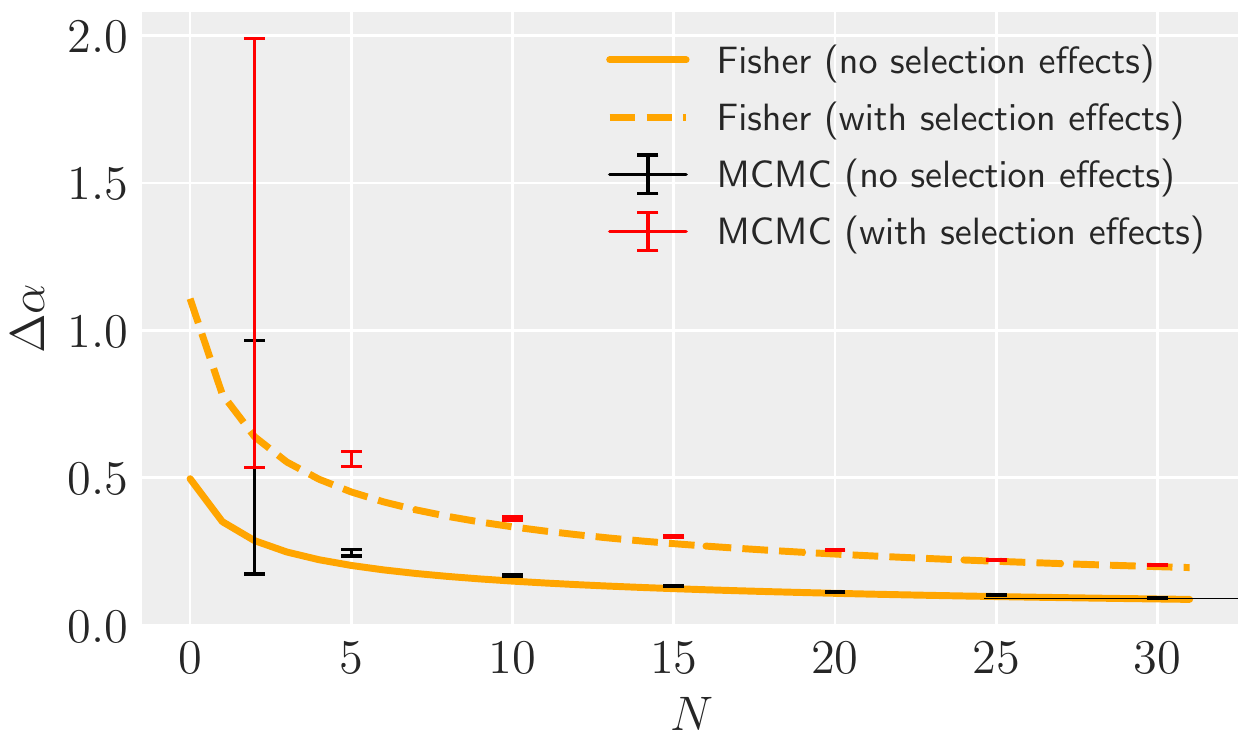}
\caption{Fisher predictions for the width of the spectral index $\alpha$ as a function of the number of observed events $N$, compared to the range of measured uncertainties obtained over a set of MCMC runs (red and black points with error bars). %The prediction refers to the case in which there are no selection effects, while t
The Fisher matrix prediction is approximated with Eq. \eqref{eq:Fisher_GWlike_example_analytical}. The agreement between the Fisher matrix predictions and the MCMC results is very good, especially if the number of events is increased beyond $N\sim 10$}. 
\label{fig:Da_vs_N}
\end{figure}

% We find this term to be enough to reproduce the MCMC results at a similar level of precision shown in Fig \ref{fig:GWlike_example}. 
We expect that the accuracy of the population Fisher matrix prediction should improve as the number of observations included in the analysis, $N_\text{tot}$, increases. 
%The relative ease in obtaining the above equation, and the fact that it leads to results that are consistent with MCMC results both independent and in the literature, can be exploited to reliably assess whether our Fisher predictions agree well with MCMC analyses as  is varied. In 
We assess this by comparing the result of MCMC analyses of data sets with increasing numbers of observations to the population Fisher matrix. We do this first for the case without selection effects, so that we can make use of the analytical prediction given above. These results are shown in Fig. \ref{fig:Da_vs_N} as $N_\text{tot}$ is varied from 2 to 30 events. For each $N_\text{tot}$, we repeat the MCMC analysis several times to allow us to estimate the variance in the posterior width between different runs. This variance is larger when there are fewer events, as expected, and so more MCMC analyses were performed for lower $N_\text{tot}$'s to ensure the variance was accurately characterised. %(up to a maximum of 30 for $N_\text{tot}=2$), while less are needed as the errors shrink at higher $N_\text{tot}$'s.
We find that for $N_\text{tot}\gtrapprox 10$ the (simplified) Fisher widths agree well with the numerical simulations. For $N_\text{tot}< 10$, the differences are progressively more pronounced, but the variance in the MCMC widths also increases, and the Fisher matrix prediction is usually within the range spanned by the MCMC runs. The agreement becomes worse for very small numbers of observations, consistent with the expectation that this is an approximation valid in the limit of large $N_\text{tot}$. Finally, we check whether a similar level of agreement is seen in the case when the observations are subject to selection effects. For this, we obtain the population Fisher matrix prediction (dashed orange line) by rescaling the $\Delta\alpha$ prediction in Fig. \ref{fig:GWlike_example} by a factor $N_\text{det}^{-1/2}$. These results are also shown in Fig. \ref{fig:GWlike_example}. We see a similar trend --- the population Fisher matrix is very accurate for $N_\text{tot} > 10$, but the accuracy diminishes for very small numbers of observations, as expected.

\subsection{More realistic scenario: SNR distribution in the population}
\label{sec:GWlike_real}
As a final example we will now make the previous scenario slightly more realistic by adding an additional property to each source, the signal-to-noise ratio. This example demonstrates how to compute the population Fisher matrix in a more realistic setting in which the measurement uncertainties depend on the source parameters, and with a more realistic model of selection effects.

For this example we assume that individual events are characterised by two parameters --- a mass, $M$, drawn from the same power law population used in the previous example, and a signal-to-noise ratio (SNR), $\rho$. We assume that $\rho$ scales with the inverse of distance and that distances are uniform in Euclidean volume, so we have $p(\rho) \propto \rho^{-4}$. We additionally assume that the SNR of a source at a particular distance is proportional to the mass. These assumptions are encoded in the SNR distribution
\begin{align}
P(\rho < P) &= \left\{\begin{array}{ll} 1- \left( \frac{M}{d_{\rm max} P}\right)^3&P > \frac{M}{d_{\rm max}} \\ 0&\mbox{otherwise}\end{array}\right. .
\end{align}
The parameter $d_{\rm max}$ represents a maximum distance for sources in the population and sets a lower limit on the SNR distribution which avoids divergences. In practice we choose $d_{\rm max} \gg M_{\rm max}/\rho_{\rm th}$, where $\rho_{\rm th}$ is the SNR threshold for detection, so that the exact choice of $d_{\rm max}$ does not influence parameter estimation. We note that the assumption that the SNR distribution is biased toward higher values for higher mass systems is not a particularly good model for EMRIs. It would be more appropriate for massive black hole binary systems, but even then the shape of the LISA sensitivity curve is such that this would only apply in a certain range of masses. We choose to make this assumption since we want to demonstrate that the population Fisher matrix works even when there are more complicated interactions between the source parameters, including parameters whose distribution is independent of the population parameters.

We assume that a GW observation consists of a noisy measurement, $\hat\rho$, of $\rho$, and a noisy measurement, $\hat{\rho M}$, of $\rho M$, so that the GW likelihood is
\begin{align}
p(\mathbf{d} = (\hat\rho,\hat{\rho M}) | \vec\theta) &= \frac{1}{2\pi \sigma_\rho \sigma_M} \exp\left[-\frac{(\hat\rho-\rho)^2}{2 \sigma_\rho^2}\right] \nonumber \\
&\hspace{2cm} \times \exp\left[-\frac{(\hat{\rho M}-\rho M)^2}{2 \sigma_M^2}\right].
\end{align}
We fix $\sigma_\rho=1$, which follows from the definition of $\rho$. %, and take $\sigma_M = 10 M_\odot$. 
We assume that selection is based on $\hat\rho$ only, with events with $\hat\rho > \rho_{\rm th}$ being deemed detectable. We use $\rho_{\rm th} = 10$ in this example. With this likelihood, the individual source Fisher matrix is
\begin{equation}
    \Gamma_{ij} = \left( \begin{array}{cc} 1+{M^2}/{\sigma_M^2}&{\rho M}/{\sigma_M^2}\\{\rho M}/{\sigma_M^2}&{\rho^2}/{\sigma_M^2} \end{array}\right).
    \label{eq:gamma_theta_GWlike}
\end{equation}
We see that the measurement uncertainties vary from event to event, with uncertainties in mass scaling like $1/\rho$, as desired~\footnote{We could have achieved the same result by assuming that the GW data comprises a measurement of $\rho$ and of $M$, with independent Gaussian errors with variances $\sigma_\rho = 1$ and $\sigma_M/\rho$ respectively. However, this model can not be put into the standard GW likelihood form, Eq.~(\ref{eq:innprod}), which assumes the noise variances are parameter independent. This alternative form of the model can be analysed using the generalised formalism described in Appendix~\ref{app:GenLike}, but we wanted the model to be of the standard GW form. The analysed model is equivalent to setting $\mathbf{h} = (\rho, \rho M)$ and $S_h(f) = (\sigma_\rho^2, \sigma^2_M)$ in Eq.~(\ref{eq:innprod}).}

\begin{figure*}
\centering
\includegraphics[width=0.8\textwidth]{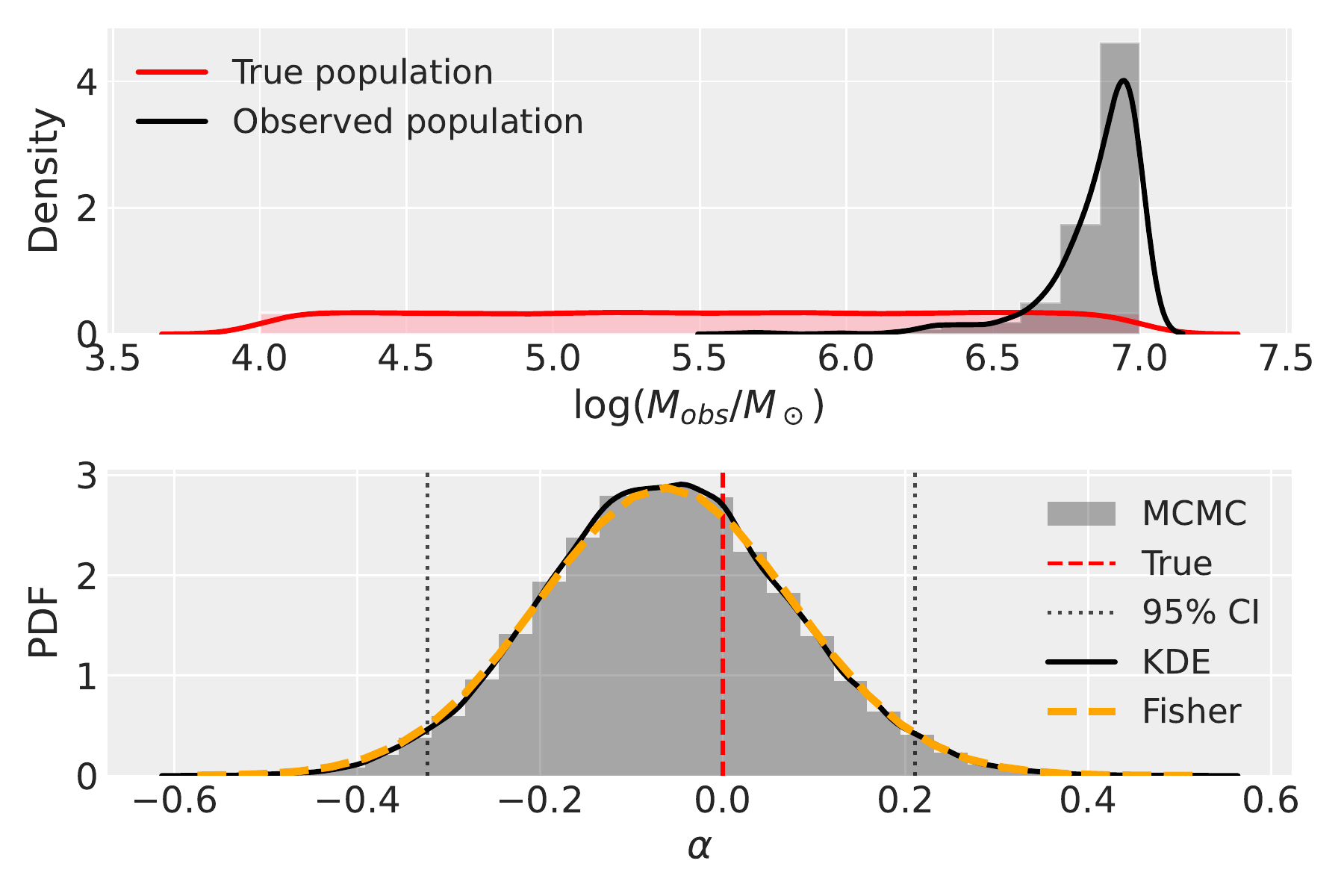}
\caption{As Figure~\ref{fig:GWlike_example} but for the slightly more realistic GW-like population model. \emph{Top panel:} distribution of masses for the underlying true population of Sec. \ref{sec:EMRI-model}  (red) and for the observed population (dark gray). 
\emph{Bottom panel:} MCMC posterior distribution for the spectral index describing the mass distribution. % in the upper panel. 
The histogram and KDE are compared against the Fisher estimate obtained as described in the text.}
\label{fig:GWlike_example_real}
\end{figure*}

We simulate observations of a population of events with true slope $\alpha = 0$ and $N_{\rm obs}=499$ observed events, by drawing $500/P_{\rm det}(\alpha) \approx 28370$ trial systems from the underlying population. We analyse these events using MCMC and compare to the predictions of the population Fisher matrix. Details of how the latter is calculated can be found in Appendix~\ref{app:GWlike_real_FM}. For this example, in the limit $\sigma_M \rightarrow 0$, the matrices $\Gamma_X$ for $X=$ II, $\cdots$, V do not vanish, because we have fixed $\sigma_\rho = 1$. However, for a reasonable choice of $\sigma_M = 10 M_\odot$, we find that the contributions from these matrices are again sub-dominant to the contribution from $\Gamma_\text{I}$, making only a $\sim 5\%$ change to the prediction for the uncertainty on $\alpha$. This is true for a wide range of choices of $\sigma_M$ up to at least $10^4M_\odot$. A comparison of the MCMC results and the population Fisher matrix prediction is shown in Figure~\ref{fig:GWlike_example_real}, demonstrating once again very precise agreement.

This example has demonstrated that the population Fisher matrix gives accurate predictions even when using a more complicated model that includes non-trivial selection effects and hetero-scedastic measurement errors. This model is a more realistic representation to a GW observation scenario, but the result cannot be directly compared to the results presented in GTV, because the assumed SNR distribution is different. Here we have assumed that sources are distributed uniformly in Euclidean space, following an $\rho^{-4}$ distribution, with SNRs additionally increased in proportion to $M$. In GTV, sources were distributed based on a computation of SNR that included the impact of other parameters, in particular time to coalescence. Nearby sources generate enough SNR that they can be observed several years before merger, which enhances the rate of nearby events and partially compensates for the fact that there are a larger number of systems further away. For this reason, the fact that we find a distribution that is approximately a factor of $3$ broader than GTV is not a cause of concern. Indeed, it is remarkable that the agreement was so close for the simpler example considered in section~\ref{sec:GWlike_simp}.

\begin{figure}
\centering
\includegraphics[width=\linewidth]{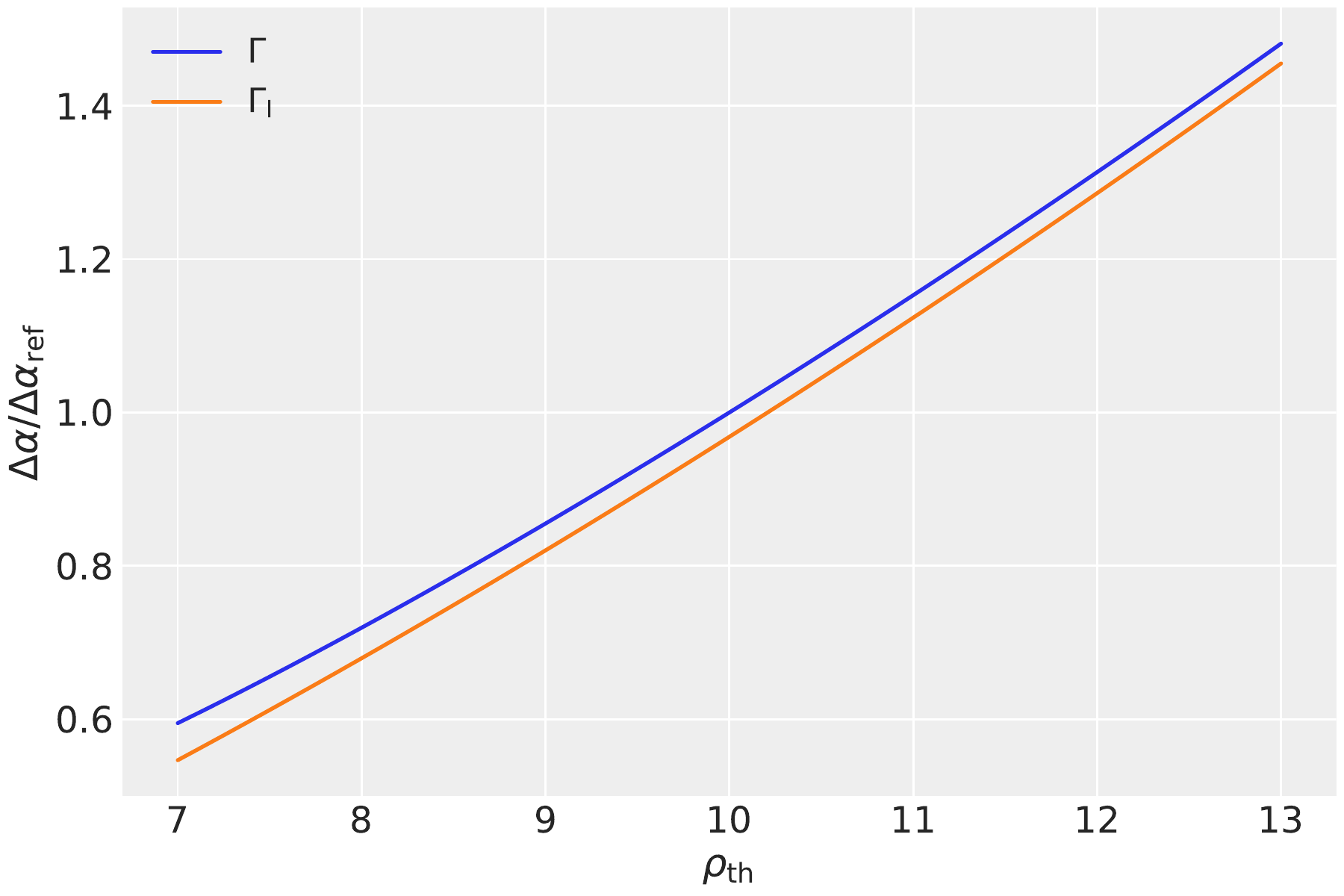}
\caption{Precision of measurement of the slope of the black hole population distribution obtained in a fixed observation time, as a function of the threshold needed for detection, $\rho_{\rm th}$. The blue curve shows results using the full population Fisher matrix, while the yellow curve shows results based on $\Gamma_\text{I}$ only. Both curves are expressed as ratios relative to the precision estimated from the full Fisher matrix with $\rho_{\rm th}=10$.}
\label{fig:prec_v_thresh}
\end{figure}

We conclude this section by using the population Fisher matrix to explore the impact of the detection threshold, $\rho_{\rm th}$, on the precision of inference of the population parameters for this simple model. There are two effects of changing the threshold. One effect is that the population Fisher matrix changes. This represents the average uncertainty over detected events and so the elements of $\Gamma_\lambda$ tend to become smaller as the threshold is decreased, corresponding to a worse constraint per event. This is because lower SNR events tend to provide less precise parameter estimates. The second effect is that the detection probability, $P_{\rm det}(\vec\lambda=\{\alpha\})$, changes, increasing as $\rho_{\rm th}$ decreases. It is therefore useful to consider the quantity $\Gamma^{-1}_{\alpha\alpha}/\sqrt{P_{\rm det}(\alpha)}$, %1/\sqrt{P_{\rm det}(\alpha) \Gamma_{\alpha\alpha}}$, 
where $\Gamma^{-1}_{\alpha\alpha}$ is the diagonal element of the inverse of the population Fisher matrix. This quantity is a measure of the precision on $\alpha$ that could be obtained in a fixed amount of observation time. We show this quantity, relative to its value for the reference threshold, $\rho_{\rm th}=10$, in Figure~\ref{fig:prec_v_thresh}. We see that the precision improves as the threshold is lowered, indicating that the increase in the number of events outweighs the decrease in the average precision per event. In practice there will be some limit to how much we can lower the threshold, beyond which we can no longer confidently identify events, or run into limitations on computational power, but within those constraints these results suggest we should lower the threshold as much as possible. In general it is at low thresholds that the approximation that the individual events can be well represented by the Fisher matrix will become less valid. This is not fully captured here because of the simplified assumption of the Gaussian likelihood. Moreover, a lot of the trend is captured by $\Gamma_\text{I}$, which is independent of that approximation. The yellow line in Figure~\ref{fig:prec_v_thresh} shows the precision estimated from $\Gamma_\text{I}$ alone, again expressed relative to the value estimated from the full Fisher matrix with $\rho_{\rm th}=10$. We see that the trend is similar. There is a slightly bigger difference between the full and approximate Fisher matrices for the lowest values of $\rho_{\rm th}$, but for all thresholds considered the approximate Fisher matrix gives a good indication of the achievable precision.

The trend in Figure~\ref{fig:prec_v_thresh} is specific to the simple model considered here and the behaviour will be different in other contexts. However, this exercise illustrates the usefulness of using the population Fisher matrix to quickly assess the impact of different assumptions on the accuracy of inference. We note, however, that when using it to assess the contribution from low-SNR events it is important to check the accuracy of the approximation in that regime, as discussed in Section~\ref{sec:validity}, to ensure that the conclusions are robust.

%%%%%%%%%%%%%%%%%%%%%%%%%%%%%%%%%%%%%%%%%%%%%%%%%%
\section{Conclusions} 
%%%%%%%%%%%%%%%%%%%%%%%%%%%%%%%%%%%%%%%%%%%%%%%%%%
\label{sec:conclusions}

The Fisher information matrix is a valuable tool for estimating the precision attainable in parameter inference, especially in contexts where the cost of doing full posterior estimation via Bayesian sampling is highly expensive~\citep{Vallisneri:2007ev}. The Fisher matrix has been widely used in GW analyses to make forecasts for the precision with which the \emph{source} parameters describing individual GW signals can be estimated by current and future detectors. In this paper we have extended the Fisher matrix concept to the estimation of the parameters characterising the \emph{population} from which a set of observed sources is drawn. Our result was derived from Eq.~\eqref{eq:popfishgen}, which is the most general definition for the population Fisher matrix. {We obtained Eq.~\eqref{eq:gammalambda}, which is valid under the assumption that individual events are observed with high enough signal-to-noise ratio that the measurement uncertainties can be well approximated by the linear signal approximation. We also identified the part of population Fisher matrix that is independent of measurement uncertainties, given by $\Gamma_\text{I}$, which is even simpler to evaluate and provides a good approximation when the individual event measurement uncertainties are much smaller than the scale on which the population varies.} %the signal-to-noise ratio of the individual events making up the population is high.
We have tested this result both analytically and against numerical Monte-Carlo results for a reference Gaussian model (Sec. \ref{sec:gaussian-gaussian}) and for a more GW-like scenario (Sec. \ref{sec:EMRI-model}), in which we are use GW events to estimate the slope of a power-law population. We find that Eq. \eqref{eq:gammalambda} is generally in very good agreement with the numerical results, for a sufficiently large number of observations. In this case sufficiently large was only $O(10)$. Results for the power-law population case can be compared to previous results in the literature \citep{Gair:2010yu}, and are found to be in very good agreement, {despite very different assumptions}. We conclude that we can reproduce the results of extensive sets of computationally expensive MCMC simulations much more cheaply, while also {correctly including selection effects}. In addition, we found that in the GW-like example the {measurement-error-independent part of the population Fisher matrix}, $(\Gamma_\text{I})_{ij}$ in Eq. \eqref{eq:gammalambda}, is sufficient to accurately reproduce the precision estimated from the full Fisher matrix. %in both of the examples considered. This is most likely due to the fact that 
This is because the noise-induced uncertainty in the parameter measurements of each individual event, $\sigma$, is sufficiently smaller than the scale on which the population model varies, that measurement errors are essentially ignorable. This result could be useful to further reduce the computational cost of computing the population Fisher matrix in other contexts.

{We note that these results are based on the approximation that individual event measurements are well represented by the individual event Fisher matrix. There will be contexts in which this is not true, but measurement uncertainties are important so $\Gamma_\text{I}$ is not dominant. We provided a criterion in Section~\ref{sec:validity} that can be used to evaluate the validity of the approximation. When the approximation is not valid, Eq.~(\ref{eq:gammalambda}) should still provide a rough estimate of the precision of inference, or the threshold can be increased such that the approximations are valid and a conservative estimate of precision obtained in this way.}

The results presented in this paper represent the first attempt at describing population inference within a Fisher formalism for generic population models, and with a likelihood that takes into account selection effects in the way of \citep{Mandel:2018mve}. The formalism developed here can be used to obtain forecasts for the precision of population analyses with future ground-based and spaceborne detectors, which are expected to detect many thousands (or even millions) of signals. Obtaining such population inference forecasts in specific contexts of relevance to current and future observations is one possible future direction for the present project. Finally, it would be interesting to generalize the results of \citep{Cutler:2007mi} and \citep{Antonelli:2021vwg} to assess inference biases on population parameters from waveform modelling errors or confusion noise.
\newline

\noindent 
\textit{Acknowledgments.} 
A.A is supported by NSF Grants No. AST-2006538, PHY-2207502, PHY-090003 and PHY20043, and NASA Grants No. 19-ATP19-0051, 20-LPS20- 0011 and 21-ATP21-0010.
\newline

\noindent 
\textit{Data Availability Statement.}
The results in this paper use \texttt{numpy} \citep{harris2020array}, \texttt{matplotlib} \citep{Hunter:2007}, \texttt{seaborn} \citep{Waskom2021}, \texttt{emcee} \citep{Foreman-Mackey:2012any}, \texttt{arviz} \citep{arviz_2019}. The results can be fully reproduced with codes made publicly available at \url{https://github.com/aantonelli94/PopFisher}.

\bibliographystyle{mn2e}
\bibliography{refs}

\begin{thebibliography}{37}
\expandafter\ifx\csname natexlab\endcsname\relax\def\natexlab#1{#1}\fi

\bibitem[{Abbott {et~al}\mbox{.}(2019)Abbott {et~al.}}]{LIGOScientific:2018mvr}
Abbott B.~P., {et~al.}, 2019, Phys. Rev. X, 9, 031040

\bibitem[{Abbott {et~al}\mbox{.}(2021{\natexlab{a}})Abbott
  {et~al.}}]{LIGOScientific:2021aug}
Abbott R., {et~al.}, 2021{\natexlab{a}}

\bibitem[{Abbott {et~al}\mbox{.}(2021{\natexlab{b}})Abbott
  {et~al.}}]{LIGOScientific:2021usb}
Abbott R., {et~al.}, 2021{\natexlab{b}}

\bibitem[{Abbott {et~al}\mbox{.}(2021{\natexlab{c}})Abbott
  {et~al.}}]{LIGOScientific:2021djp}
Abbott R., {et~al.}, 2021{\natexlab{c}}

\bibitem[{Abbott {et~al}\mbox{.}(2021{\natexlab{d}})Abbott
  {et~al.}}]{LIGOScientific:2021psn}
Abbott R., {et~al.}, 2021{\natexlab{d}}

\bibitem[{Amaro-Seoane {et~al}\mbox{.}(2007)Amaro-Seoane, Gair, Freitag,
  Coleman~Miller, Mandel, Cutler, \& Babak}]{Amaro-Seoane:2007osp}
Amaro-Seoane P., Gair J.~R., Freitag M., Coleman~Miller M., Mandel I., Cutler
  C.~J., Babak S., 2007, Class. Quant. Grav., 24, R113

\bibitem[{Amaro-Seoane {et~al}\mbox{.}(2017)Amaro-Seoane
  {et~al.}}]{Audley:2017drz}
Amaro-Seoane P., {et~al.}, 2017

\bibitem[{Antonelli, Burke \& Gair(2021)Antonelli, Burke, \&
  Gair}]{Antonelli:2021vwg}
Antonelli A., Burke O., Gair J.~R., 2021, Mon. Not. Roy. Astron. Soc., 507,
  5069

\bibitem[{Babak {et~al}\mbox{.}(2017)Babak, Gair, Sesana, Barausse, Sopuerta,
  Berry, Berti, Amaro-Seoane, Petiteau, \& Klein}]{Babak:2017tow}
Babak S. {et~al.}, 2017, Phys. Rev. D, 95, 103012

\bibitem[{Barack(2009)}]{Barack:2009ux}
Barack L., 2009, Class. Quant. Grav., 26, 213001

\bibitem[{Barausse, Cardoso \& Pani(2014)Barausse, Cardoso, \&
  Pani}]{Barausse:2014tra}
Barausse E., Cardoso V., Pani P., 2014, Phys. Rev. D, 89, 104059

\bibitem[{Bonetti \& Sesana(2020)}]{Bonetti:2020jku}
Bonetti M., Sesana A., 2020, Phys. Rev. D, 102, 103023

\bibitem[{Borhanian(2021)}]{Borhanian:2020ypi}
Borhanian S., 2021, Class. Quant. Grav., 38, 175014

\bibitem[{Cutler \& Vallisneri(2007)}]{Cutler:2007mi}
Cutler C., Vallisneri M., 2007, Phys. Rev. D, 76, 104018

\bibitem[{Fishbach {et~al}\mbox{.}(2021)Fishbach, Doctor, Callister, Edelman,
  Ye, Essick, Farr, Farr, \& Holz}]{Fishbach:2021yvy}
Fishbach M. {et~al.}, 2021, Astrophys. J., 912, 98

\bibitem[{Foreman-Mackey {et~al}\mbox{.}(2013)Foreman-Mackey, Hogg, Lang, \&
  Goodman}]{Foreman-Mackey:2012any}
Foreman-Mackey D., Hogg D.~W., Lang D., Goodman J., 2013, Publ. Astron. Soc.
  Pac., 125, 306

\bibitem[{Gair {et~al}\mbox{.}(2011)Gair, Sesana, Berti, \&
  Volonteri}]{Gair:2010bx}
Gair J.~R., Sesana A., Berti E., Volonteri M., 2011, Class. Quant. Grav., 28,
  094018

\bibitem[{Gair, Tang \& Volonteri(2010)Gair, Tang, \& Volonteri}]{Gair:2010yu}
Gair J.~R., Tang C., Volonteri M., 2010, Phys. Rev. D, 81, 104014

\bibitem[{Gair {et~al}\mbox{.}(2013)Gair, Vallisneri, Larson, \&
  Baker}]{Gair:2012nm}
Gair J.~R., Vallisneri M., Larson S.~L., Baker J.~G., 2013, Living Rev. Rel.,
  16, 7

\bibitem[{Harms {et~al}\mbox{.}(2022)Harms, Dupletsa, Banerjee, Branchesi,
  Goncharov, Maselli, Oliveira, Ronchini, \& Tissino}]{Harms:2022ymm}
Harms J. {et~al.}, 2022

\bibitem[{Harris {et~al}\mbox{.}(2020)Harris, Millman, van~der Walt, Gommers,
  Virtanen, Cournapeau, Wieser, Taylor, Berg, Smith, Kern, Picus, Hoyer, van
  Kerkwijk, Brett, Haldane, del R{\'{i}}o, Wiebe, Peterson,
  G{\'{e}}rard-Marchant, Sheppard, Reddy, Weckesser, Abbasi, Gohlke, \&
  Oliphant}]{harris2020array}
Harris C.~R. {et~al.}, 2020, Nature, 585, 357

\bibitem[{Hunter(2007)}]{Hunter:2007}
Hunter J.~D., 2007, Computing in Science \& Engineering, 9, 90

\bibitem[{Kumar {et~al}\mbox{.}(2019)Kumar, Carroll, Hartikainen, \&
  Martin}]{arviz_2019}
Kumar R., Carroll C., Hartikainen A., Martin O., 2019, Journal of Open Source
  Software, 4, 1143

\bibitem[{Mancarella, Genoud-Prachex \& Maggiore(2022)Mancarella,
  Genoud-Prachex, \& Maggiore}]{Mancarella:2021ecn}
Mancarella M., Genoud-Prachex E., Maggiore M., 2022, Phys. Rev. D, 105, 064030

\bibitem[{Mandel, Farr \& Gair(2019)Mandel, Farr, \& Gair}]{Mandel:2018mve}
Mandel I., Farr W.~M., Gair J.~R., 2019, Mon. Not. Roy. Astron. Soc., 486, 1086

\bibitem[{Mastrogiovanni {et~al}\mbox{.}(2021)Mastrogiovanni, Leyde,
  Karathanasis, Chassande-Mottin, Steer, Gair, Ghosh, Gray, Mukherjee, \&
  Rinaldi}]{Mastrogiovanni:2021wsd}
Mastrogiovanni S. {et~al.}, 2021, Phys. Rev. D, 104, 062009

\bibitem[{Mould {et~al}\mbox{.}(2022)Mould, Gerosa, Broekgaarden, \&
  Steinle}]{Mould:2022xeu}
Mould M., Gerosa D., Broekgaarden F.~S., Steinle N., 2022

\bibitem[{Mukherjee {et~al}\mbox{.}(2022)Mukherjee, Krolewski, Wandelt, \&
  Silk}]{Mukherjee:2022afz}
Mukherjee S., Krolewski A., Wandelt B.~D., Silk J., 2022

\bibitem[{Punturo {et~al}\mbox{.}(2010)Punturo {et~al.}}]{Punturo:2010zz}
Punturo M., {et~al.}, 2010, Class. Quant. Grav., 27, 194002

\bibitem[{Reitze {et~al}\mbox{.}(2019)Reitze {et~al.}}]{Reitze:2019iox}
Reitze D., {et~al.}, 2019, Bull. Am. Astron. Soc., 51, 035

\bibitem[{Rodriguez {et~al}\mbox{.}(2020)Rodriguez
  {et~al.}}]{Rodriguez:2020viw}
Rodriguez C.~L., {et~al.}, 2020, Astrophys. J. Lett., 896, L10

\bibitem[{Sesana {et~al}\mbox{.}(2011)Sesana, Gair, Berti, \&
  Volonteri}]{Sesana:2010wy}
Sesana A., Gair J., Berti E., Volonteri M., 2011, Phys. Rev. D, 83, 044036

\bibitem[{Taylor \& Gerosa(2018)}]{Taylor:2018iat}
Taylor S.~R., Gerosa D., 2018, Phys. Rev. D, 98, 083017

\bibitem[{Vallisneri(2008)}]{Vallisneri:2007ev}
Vallisneri M., 2008, Phys. Rev. D, 77, 042001

\bibitem[{Vitale, Biscoveanu \& Talbot(2022)Vitale, Biscoveanu, \&
  Talbot}]{Vitale:2022pmu}
Vitale S., Biscoveanu S., Talbot C., 2022

\bibitem[{Vitale {et~al}\mbox{.}(2020)Vitale, Gerosa, Farr, \&
  Taylor}]{Vitale:2020aaz}
Vitale S., Gerosa D., Farr W.~M., Taylor S.~R., 2020

\bibitem[{Waskom(2021)}]{Waskom2021}
Waskom M.~L., 2021, Journal of Open Source Software, 6, 3021

\end{thebibliography}

\appendix

\section{Dealing with rates}
\label{app:rates}
%\jg{Andrea, Riccardo: are you enlightened??} \andrea{Seems very nicely explained, but I can't say I understand it 100 percent.}
The likelihood in Eq.~\eqref{eq:likewithsel} assumes that the number of events that are detected conveys no information about the population. Relaxing this assumption the joint likelihood takes the alternative form
\begin{align}
    p(\{ \mathbf{d}_i \} | \vec\lambda, R) \propto \left[\prod_{i=1}^n p_{\rm full}(\mathbf{d}_i | \vec\lambda) \right] R^n \exp[-R P_{\rm det} (\vec\lambda)], \label{eq:likewithrate}
\end{align}
in which $R$ is the rate of events occurring in the Universe over the total time data has been collected, and all other terms are as before. The derivation of this expression can be found in~\citep{Mandel:2018mve}. Imposing an (improper) scale-invariant prior on the total rate, $p(R) \propto 1/R$, and marginalising over $R$ we obtain the form of the joint likelihood used in Eq.~\eqref{eq:likewithsel} and~\eqref{eq:jointlike}. 

We denote the rate-dependent terms by
\be
p_{\rm rate} (n | R, \vec\lambda) \equiv R^n \exp[-R P_{\rm det} (\vec\lambda)].
\ee
The contribution of these terms to the joint log-likelihood is 
\be
\ln p_{\rm rate} (n | R, \vec\lambda) = n \ln R - R P_{\rm det}(\vec\lambda),
\ee
which is maximized when $R = n/P_{\rm det}(\vec\lambda)$. Expanding about this maximum-likelihood point we can write $R = n/P_{\rm det} + \delta R$ and obtain
\begin{align}
\label{eq:likewithrate_expanded}
    \ln p_{\rm rate} (n | R, \vec\lambda) &= n \ln n - n \ln P_{\rm det}(\vec\lambda) + n \ln (1 + \delta R P_{\rm det}/n) \nonumber \\
    &\hspace{2cm} -n-\delta R P_{\rm det} \nonumber \\
    &= (n \ln n - n) - n \ln P_{\rm det}(\vec\lambda) - \frac{\delta R^2 P_{\rm det}^2}{2 n} + \cdots .
\end{align}
The second term here, $-n \ln P_{\rm det}(\vec\lambda)$, is what is needed to change $p_{\rm full}(\mathbf{d}_i | \vec\lambda)$ into $p(\mathbf{d}_i | \vec\lambda)$ in the product term in Eq.~\eqref{eq:likewithrate}, reducing that to the form analysed in the main body of the paper. We deduce that the asymptotic Fisher matrix for the joint estimation of $R$ and $\vec\lambda$ is block diagonal
\be
\Gamma = \left( \begin{array}{cc} \Gamma_\lambda&0 \\0&\Gamma_R \end{array} \right)
\ee
with the shape parameter block, $\Gamma_\lambda$, as before and the rate precision given by the inverse of $\Gamma_R = P_{\rm det}^2/2n$, where this result can be obtained directly from the coefficient of $\delta R^2$ in equation \eqref{eq:likewithrate_expanded}. We conclude that the precision with which the shape parameters can be determined does not depend on which particular form of the likelihood is being used. This makes sense since we know that the two forms are equivalent for a particular choice of rate prior, and we expect results to be asymptotically independent of the initial prior choice. We note also that the precision with which the rate of observed events, $R P_{\rm det}$, can be measured is $\sqrt{n}$, consistent with the expected uncertainty in the estimation of the rate of a Poisson process.

To conclude this section, we note that in the above we have been assuming that the rate parameter $R$ is an additional parameter of the model, separate to the parameters $\vec\lambda$ that characterise the shape of the population distribution. If instead both $R$ and $\vec\lambda$ are functions of another set of population parameters, $\vec\mu$, we can use the linear signal approximation to change variables and obtain the usual result that the Fisher matrix for the $\vec\mu$ parameters is
\be
(\Gamma_\mu)_{cd} = (\Gamma_\lambda)_{ab} \frac{\partial \lambda^a}{\partial \mu^c} \frac{\partial \lambda^b}{\partial \mu^d} + \Gamma_R  \frac{\partial R}{\partial \mu^c} \frac{\partial R}{\partial \mu^d}.
\ee
In this case, the result is different to what would be obtained by transforming the Fisher matrix that ignores the rate, which would be the first term only. This reflects the fact that if $R$ also depends on $\vec\mu$, the measurement of $R$ provides additional information that can help to improve the estimation of the $\vec\mu$ parameters.

\section{The asymptotic behaviour of the Fisher matrix}
\label{app:popFMcorrect}
 In this section we will rederive the expression for the Fisher matrix given in the main body of the paper by directly expanding the posterior distribution. In doing so we will derive the form and scaling of the leading corrections to the Fisher matrix approximation. We will proceed by computing the posterior mode, mean and variance in the limit $n \gg 1$. These are all random variables, since they depend on the particular realisation of the data that is being analysed, and so we can characterise them by their expectation value and variance. We will show how to compute the first two terms in a large-$n$ expansion of both the mean and variance for all three posterior summary statistics, and give the result explicitly for the posterior mode. We note that similar results for corrections to the individual event Fisher matrix were given in~\citep{Vallisneri:2007ev}, but those relied on the assumption of a Gaussian likelihood which permits simplifications. The expansion presented here is valid for any population-level likelihood, $p(\mathbf{d}|\vec\lambda)$.
We write
\begin{equation}
\hat{\mu}(n,\{\mathbf{d}_i\}, \vec\lambda) = -\frac{1}{n} \sum_{i=1}^n \ln p(\mathbf{d}_i | \vec\lambda).
\end{equation}
such that the posterior distribution is proportional to $\exp[-n \hat\mu(n,\{\mathbf{d}_i\}, \vec\lambda)  + \ln \pi(\vec\lambda)]$. We use  $\vec{\beta}(n,\{\mathbf{d}_i\}, \vec\lambda_t)$ to denote the solution to
\begin{equation}
U_i +U^\pi_i + (V_{ij} +V_{ij}^\pi)\beta^j=0
\end{equation}
where 
\begin{align}
U_i &= \left(\frac{\partial \hat\mu}{\partial \lambda^i}\right)_{|\vec\lambda_t}, \qquad U^\pi_i = -\frac{1}{n}\left(\frac{\partial \ln\pi}{\partial \lambda^i}\right)_{|\vec\lambda_t} \nonumber \\
V_{ij} &= \left( \frac{\partial^2 \hat\mu}{\partial \lambda^i \partial\lambda^j}\right)_{|\vec\lambda_t}, \qquad V^\pi_{ij} = -\frac{1}{n}\left(\frac{\partial^2 \ln\pi}{\partial \lambda^i \partial\lambda^j}\right)_{|\vec\lambda_t}.
\end{align}
The quantity $\hat\mu$, and its derivatives, are averages of a set of independent identically distributed (IID) random variables and so have predictable scalings. The expectation value is $O(1)$, covariances are $O(1/n)$, three and four point functions are $O(1/n^2)$, and so on. In this case $\mathbb{E}|[\mathbb{U}]=0$, as shown in Eq.~\eqref{eq:unbiasproof} in the main body of the paper. This facilitates obtaining a solution for $\vec\beta$ perturbatively
\begin{align}
    \vec\beta &=  \vec\beta_{\frac{1}{2}} +  \vec\beta_{1} +  \vec\beta_{\frac{3}{2}} + \cdots \nonumber \\
    \mbox{where } \vec\beta_{\frac{1}{2}} &=(\mathbb{E}[\mathbf{V}])^{-1} \mathbf{U} \nonumber \\
     \vec\beta_{1}&= (\mathbb{E}[\mathbf{V}])^{-1} \left[(\mathbf{V}-\mathbb{E}[\mathbf{V}])  \vec\beta_{\frac{1}{2}}+ \mathbf{U}^\pi\right] \nonumber \\
     \vec\beta_{\frac{3}{2}}&= (\mathbb{E}[\mathbf{V}])^{-1} \left[ (\mathbf{V}-\mathbb{E}[\mathbf{V}])  \vec\beta_{1}+\mathbf{V}^\pi  \vec\beta_{\frac{1}{2}}\right],
\end{align}
in which $\vec\beta_{k} \sim n^{-k}$.

The $\vec\beta_{k}$'s are random variables, but we can compute their means 
\begin{align}
\mathbb{E}\left[ \vec\beta_{\frac{1}{2}}\right] &= 0 \nonumber \\
\mathbb{E}\left[ \beta^i_{1}\right] &=  (\mathbb{E}[\mathbf{V}])^{-1}_{ij} \left[ (  \mathbb{E}[\mathbf{V}])^{-1}_{kl} \Sigma^{VU}_{jkl}+\mathbf{U}_j^\pi \right] \nonumber \\
\mathbb{E}\left[ \beta^i_{\frac{3}{2}}\right] &= (\mathbb{E}[\mathbf{V}])^{-1}_{ij} (\mathbb{E}[\mathbf{V}])^{-1}_{kl} (\mathbb{E}[\mathbf{V}])^{-1}_{mn} \Sigma^{VVU}_{jklmn} \nonumber \\
%\Sigma^{VU}_{ijk} &= \mathbb{E}\left[(\mathbf{V}-\mathbb{E}[\mathbf{V}])_{ij} \mathbf{U}_k\right] \nonumber \\
%&= \frac{1}{n} \int \left( \frac{\partial^2 \ln p(\mathbf{d} | \vec\lambda)}{\partial\lambda^i\partial\lambda^j} \right) \left( \frac{\partial \ln p(\mathbf{d} | \vec\lambda)}{\partial\lambda^k} \right) p(\mathbf{d} | \vec\lambda_t) {\rm d}\mathbf{d} \nonumber \\
\Sigma^{VU}_{ijk} &= \mathbb{E}\left[(V_{ij}-\mathbb{E}[V_{ij}]) {U}_k\right] \nonumber \\
&= \frac{1}{n} \int l_{,ij}(\mathbf{d}) l_{,k}(\mathbf{d}) p(\mathbf{d} | \vec\lambda_t) {\rm d}\mathbf{d} \nonumber \\
\Sigma^{VVU}_{ijklm} &= \mathbb{E}\left[({V}_{ij}-\mathbb{E}[{V}_{ij}]) ({V}_{kl}-\mathbb{E}[{V}_{kl}]) {U}_m\right] \nonumber \\
&= \frac{1}{n^2} \int l_{,ij}(\mathbf{d}) l_{,kl}(\mathbf{d}) l_{,m}(\mathbf{d}) \, p(\mathbf{d} | \vec\lambda_t) {\rm d}\mathbf{d} \nonumber \\
& \hspace{2cm} -\mathbb{E}[{V}_{ij}] \Sigma^{VU}_{klm} -\mathbb{E}[{V}_{kl}] \Sigma^{VU}_{ijm} 
%\Sigma^{VVU}_{ijklm} &= \mathbb{E}\left[(\mathbf{V}-\mathbb{E}[\mathbf{V}])_{ij} (\mathbf{V}-\mathbb{E}[\mathbf{V}])_{kl} \mathbf{U}_m\right] \nonumber \\
%&= \frac{1}{n^2} \int \left( \frac{\partial^2 \ln p(\mathbf{d} | \vec\lambda)}{\partial\lambda^i\partial\lambda^j} \right) \left( \frac{\partial^2 \ln p(\mathbf{d} | \vec\lambda)}{\partial\lambda^k\partial\lambda^l} \right)  \nonumber \\
%&\hspace{1.5cm} \times\left( \frac{\partial \ln p(\mathbf{d} | \vec\lambda)}{\partial\lambda^m} \right)p(\mathbf{d} | \vec\lambda_t) {\rm d}\mathbf{d} \nonumber \\
%& \hspace{2cm} -\mathbb{E}(\mathbf{V})_{ij} \Sigma^{VU}_{klm} -\mathbb{E}(\mathbf{V})_{kl} \Sigma^{VU}_{ijm} 
\end{align}
where we are using the notation $l_{,i}(\mathbf{d})$ to denote the derivative $\partial \ln p(\mathbf{d}|\vec\lambda)/\partial\lambda^i$ evaluated at $\vec\lambda=\vec\lambda_t$. Additional indices after the comma indicate further partial derivatives as usual. The fact that the first expectation value vanishes is why we have continued the expansion to three terms, allowing us to obtain the first two terms in an expansion of the mean.

We can also compute their covariances, using the usual notation cov$(a^i,b^j) = \mathbb{E}[(a^i - \mathbb{E}[a^i]) (b^j - \mathbb{E}[b^j]) ]$.
\begin{align}
\mbox{cov}(\beta_{\frac{1}{2}}^i,\beta_{\frac{1}{2}}^j) &= (\mathbb{E}[\mathbf{V}])^{-1}_{ik} (\mathbb{E}[\mathbf{V}])^{-1}_{jl} \Sigma^{UU}_{kl} \nonumber \\
\mbox{cov}(\beta_{\frac{1}{2}}^i,\beta_{1}^j)&= (\mathbb{E}[\mathbf{V}])^{-1}_{im} (\mathbb{E}[\mathbf{V}])^{-1}_{jk} (\mathbb{E}[\mathbf{V}])^{-1}_{ln} \Sigma^{VUU}_{klmn}\nonumber \\
\mbox{cov}(\beta_{1}^i,\beta_{1}^j)&= (\mathbb{E}[\mathbf{V}])^{-1}_{ik}(\mathbb{E}[\mathbf{V}])^{-1}_{lm}(\mathbb{E}[\mathbf{V}])^{-1}_{jp}(\mathbb{E}[\mathbf{V}])^{-1}_{qr} \Sigma^{VVUU}_{klpqmr} \nonumber \\
& \hspace{1cm}- (\mathbb{E}[\mathbf{V}])^{-1}_{ik}(\mathbb{E}[\mathbf{V}])^{-1}_{lm} \Sigma^{VU}_{klm}\Sigma^{VU}_{pqr}\nonumber \\
\mbox{cov}(\beta_{\frac{3}{2}}^i,\beta_{\frac{1}{2}}^j)&=(\mathbb{E}[\mathbf{V}])^{-1}_{ik}(\mathbb{E}[\mathbf{V}])^{-1}_{jq}(\mathbb{E}[\mathbf{V}])^{-1}_{lm} \nonumber \\
&\hspace{0.2cm} \times\left[ (\mathbb{E}[\mathbf{V}])^{-1}_{np} \Sigma^{VVUU}_{klmnpq} + \mathbf{U}^\pi_m \Sigma^{VU}_{klq} + \mathbf{V}^\pi_{kl} \Sigma^{VUU}_{klmq} \right]\nonumber \\
\Sigma^{UU}_{ij} &= \mathbb{E}\left[{U}_{i} {U}_j\right] \nonumber \\
%&= \frac{1}{n} \int \left( \frac{\partial \ln p(\mathbf{d} | \vec\lambda)}{\partial\lambda^i} \right) \left( \frac{\partial \ln p(\mathbf{d} | \vec\lambda)}{\partial\lambda^j} \right) p(\mathbf{d} | \vec\lambda_t) {\rm d}\mathbf{d} \nonumber \\
&= \frac{1}{n} \int l_{,i}(\mathbf{d}) l_{,j}(\mathbf{d}) \; p(\mathbf{d} | \vec\lambda_t) {\rm d}\mathbf{d} \nonumber \\
\Sigma^{VUU}_{ijkl} &= \mathbb{E}\left[({V}_{ij}-\mathbb{E}[{V}_{ij}]){U}_{k} {U}_l\right] \nonumber \\
%&= \frac{1}{n^2} \int \left( \frac{\partial^2 \ln p(\mathbf{d} | \vec\lambda)}{\partial\lambda^i \partial\lambda^j} \right) \left( \frac{\partial \ln p(\mathbf{d} | \vec\lambda)}{\partial\lambda^k} \right) \nonumber \\
%& \hspace{2cm} \left( \frac{\partial \ln p(\mathbf{d} | \vec\lambda)}{\partial\lambda^l} \right)p(\mathbf{d} | \vec\lambda_t) {\rm d}\mathbf{d} \nonumber \\
&= \frac{1}{n^2} \int l_{,ij}(\mathbf{d})l_{,k}(\mathbf{d})l_{,l}(\mathbf{d}) \; p(\mathbf{d} | \vec\lambda_t) {\rm d}\mathbf{d} \nonumber \\
&\hspace{2cm} -\mathbb{E}[{V}_{ij}] \Sigma^{UU}_{kl} \nonumber \\
\Sigma^{VVUU}_{ijklmn} &= \mathbb{E}\left[({V}_{ij}-\mathbb{E}[{V}_{ij}])({V}_{kl}-\mathbb{E}[{V}_{kl}]){U}_{m} {U}_n\right] \nonumber \\
%&= \frac{n(n-1)}{n^4}  \int \left( \frac{\partial^2 \ln p(\mathbf{d} | \vec\lambda)}{\partial\lambda^i \partial\lambda^j} \right) \left( \frac{\partial^2 \ln p(\mathbf{d} | \vec\lambda)}{\partial\lambda^k \partial\lambda^l} \right)p(\mathbf{d} | \vec\lambda_t) {\rm d}\mathbf{d}  \nonumber \\. %\nonumber \\
&= \frac{n(n-1)}{n^4} \bigg[ \int l_{,ij}(\mathbf{d})l_{,kl}(\mathbf{d})\; p(\mathbf{d} | \vec\lambda_t) {\rm d}\mathbf{d} \nonumber \\
&\hspace{1cm} \times \int l_{,m}(\mathbf{d})l_{,n}(\mathbf{d})\; p(\mathbf{d} | \vec\lambda_t) {\rm d}\mathbf{d} \nonumber \\
&\hspace{0.5cm} + \int l_{,ij}(\mathbf{d})l_{,m}(\mathbf{d})\; p(\mathbf{d} | \vec\lambda_t) {\rm d}\mathbf{d}  \nonumber \\
&\hspace{1cm} \times \int l_{,kl}(\mathbf{d})l_{,n}(\mathbf{d})\; p(\mathbf{d} | \vec\lambda_t) {\rm d}\mathbf{d} \nonumber \\
&\hspace{0.5cm} + m \leftrightarrow n \bigg]
%\int l_{,ij}(\mathbf{d})l_{,n}(\mathbf{d})\; p(\mathbf{d} | \vec\lambda_t) {\rm d}\mathbf{d}  \nonumber \\
%&\hspace{1cm} \times \left. \int l_{,kl}(\mathbf{d})l_{,m}(\mathbf{d})\; p(\mathbf{d} | \vec\lambda_t) {\rm d}\mathbf{d} \right]
\nonumber \\
&\hspace{0.2cm} + \frac{1}{n^3} \int l_{,ij}(\mathbf{d})l_{,kl}(\mathbf{d})l_{,m}(\mathbf{d})l_{,n}(\mathbf{d})\; p(\mathbf{d} | \vec\lambda_t) {\rm d}\mathbf{d} \nonumber \\
&\hspace{1.5cm} - \mathbb{E}[{V}_{ij}]\Sigma^{VUU}_{klmn} - \mathbb{E}[{V}_{kl}]\Sigma^{VUU}_{ijmn}\nonumber \\
&\hspace{2.5cm}  - \mathbb{E}[{V}_{ij}]\mathbb{E}[{V}_{kl}] \Sigma^{UU}_{mn} 
%\mbox{cov}(\beta_{1}^i,\beta_{1}^j) &= 
\end{align}
All four of these terms are needed to obtain the first two terms in the covariance of $\vec\beta$. We need only retain terms of $O(1/n^2)$ in the final expression, which means the last term in $\Sigma^{VVUU}_{ijklmn}$ can be ignored and $n(n-1)/n^4$ replaced by $n^{-2}$. The final result, keeping the first two orders is
\begin{align}
    \mbox{cov}\left(\beta^i,\beta^j\right) &= \mbox{cov}\left(\beta_{\frac{1}{2}}^i,\beta_{\frac{1}{2}}^j\right) + \mbox{cov}\left(\beta_{\frac{1}{2}}^i,\beta_1^j\right) \nonumber \\
    & \hspace{0.5cm} + \mbox{cov}\left(\beta_{\frac{1}{2}}^j,\beta_1^i\right) +  
    \mbox{cov}\left(\beta_1^i,\beta_1^j\right)  \nonumber \\
    & \hspace{0.5cm} +
    \mbox{cov}\left(\beta_{\frac{3}{2}}^i,\beta_{\frac{1}{2}}^j\right)  +
    \mbox{cov}\left(\beta_{\frac{3}{2}}^j,\beta_{\frac{1}{2}}^i\right) 
\end{align}

The posterior mode is at $\vec\lambda = \vec\lambda_t+\vec\beta+\vec\delta$. The random variable $\vec\delta$ obeys the equation
\begin{align}
    0&=\delta^j V_{ij} + \frac{1}{2} (\beta^j+\delta^j)(\beta^k+\delta^k) W_{ijk} +  V_{ij}^\pi \delta^j \nonumber \\
    & \hspace{1cm} +\frac{1}{2} (\beta^j+\delta^j)(\beta^k+\delta^k) W^\pi_{ijk} \nonumber \\
    W_{ijk}&=\left(\frac{\partial^3\hat\mu}{\partial \lambda^i \partial \lambda^j \partial \lambda^k}\right)_{|\vec\lambda_t} \nonumber \\
    W^\pi_{ijk}&=-\frac{1}{n}\left(\frac{\partial^3\ln\pi}{\partial \lambda^i \partial \lambda^j \partial \lambda^k}\right)_{|\vec\lambda_t}.
\end{align}
We can find a perturbative solution as we did for $\vec\beta$
\begin{align}
    \vec\delta&=\vec\delta_1+\vec\delta_{\frac{3}{2}} + \cdots \nonumber \\
    \mbox{where } \delta^i_1 &=\frac{1}{2} \left( \mathbb{E}[\mathbf{V}]\right)^{-1}_{ij} \mathbb{E}[{W}_{jkl}] \beta_{\frac{1}{2}}^k \beta_{\frac{1}{2}}^l \nonumber \\
    \delta^i_{\frac{3}{2}} &= \left( \mathbb{E}[\mathbf{V}]\right)^{-1}_{ij} \left[ \beta_{\frac{1}{2}}^k \beta_1^l  \mathbb{E}[{W}_{jkl}] + \beta_{\frac{1}{2}}^k \delta_1^l \mathbb{E}[{W}_{jkl}] \right. \nonumber \\
    & \hspace{1.5cm}  + \left({V}_{jk} - \mathbb{E}[{V}_{jk}]\right) \delta_1^k \nonumber \\
    & \hspace{2cm} \left.+\frac{1}{2} ({W}_{jkl}-\mathbb{E}[{W}_{jkl}]) \beta_{\frac{1}{2}}^k \beta_{\frac{1}{2}}^l\right]
\end{align}
The means and relevant covariances are
\begin{align}
    \mathbb{E}\left[ \delta_1^i\right] &= \frac{1}{2}\left( \mathbb{E}[\mathbf{V}]\right)^{-1}_{ij} \mathbb{E}[{W}_{jkl}] \;\mbox{cov}\left(\beta_{\frac{1}{2}}^k,\beta_{\frac{1}{2}}^l \right) \nonumber \\
    \mathbb{E}\left[ \delta_{\frac{3}{2}}^i\right] &= \left( \mathbb{E}[\mathbf{V}]\right)^{-1}_{ij} \left[   \mathbb{E}[{W}_{jkl}] \;\mbox{cov} \left(\beta_{\frac{1}{2}}^k, \beta_1^l\right)\right. \nonumber \\
    & \hspace{0.4cm} +\mathbb{E}[{W}_{jkl}] \; \mbox{cov}\left(\beta_{\frac{1}{2}}^k, \delta_1^l\right) \nonumber \\
    & \hspace{0.4cm}  + \frac{1}{2}  \left( \mathbb{E}[\mathbf{V}]\right)^{-1}_{kn} \mathbb{E}[{W}_{npq}] \left( \mathbb{E}[\mathbf{V}]\right)^{-1}_{pl} \left( \mathbb{E}[\mathbf{V}]\right)^{-1}_{qm} \Sigma^{VUU}_{jklm}  \nonumber \\
    & \hspace{2cm} \left.+\frac{1}{2} \left( \mathbb{E}[\mathbf{V}]\right)^{-1}_{km}
    \left( \mathbb{E}[\mathbf{V}]\right)^{-1}_{ln}
    \Sigma^{WUU}_{jklmn} \right]\nonumber \\
    \mbox{cov}\left(\delta_1^i,\delta_1^j \right) &= \frac{1}{2} \left( \mathbb{E}[\mathbf{V}]\right)^{-1}_{ik} \left( \mathbb{E}[\mathbf{V}]\right)^{-1}_{jl} \left( \mathbb{E}[\mathbf{V}]\right)^{-1}_{mx} \left( \mathbb{E}[\mathbf{V}]\right)^{-1}_{ny} \nonumber \\
    & \hspace{0.5cm} \left( \mathbb{E}[\mathbf{V}]\right)^{-1}_{pr} \left( \mathbb{E}[\mathbf{V}]\right)^{-1}_{qs}  \mathbb{E}[{W}_{kmn}]  \mathbb{E}[{W}_{lpq}] \Sigma^{UUUU}_{xyrs} \nonumber \\
    &\hspace{2cm}- \mathbb{E}[\delta_1^i] \mathbb{E}[\delta_1^i]\nonumber \\
      \mbox{cov}\left(\delta_1^i,\beta_{\frac{1}{2}}^j \right) &= \frac{1}{2} \left( \mathbb{E}[\mathbf{V}]\right)^{-1}_{ik} \left( \mathbb{E}[\mathbf{V}]\right)^{-1}_{jp} \left( \mathbb{E}[\mathbf{V}]\right)^{-1}_{lq} \left( \mathbb{E}[\mathbf{V}]\right)^{-1}_{mr}  \nonumber \\
     &\hspace{2cm} \mathbb{E}[{W}_{klm}] \Sigma^{UUU}_{pqr}\nonumber \\
      \mbox{cov}\left(\delta_1^i,\beta_1^j \right) &= \frac{1}{2} \left( \mathbb{E}[\mathbf{V}]\right)^{-1}_{ik} \left( \mathbb{E}[\mathbf{V}]\right)^{-1}_{lp}\left( \mathbb{E}[\mathbf{V}]\right)^{-1}_{mq} \left( \mathbb{E}[\mathbf{V}]\right)^{-1}_{jr} \nonumber \\
      &\hspace{1cm} \mathbb{E}[{W}_{klm}] \left(\left( \mathbb{E}[\mathbf{V}]\right)^{-1}_{sx} \Sigma^{VUUU}_{rsxpq} + {U}_r^\pi \Sigma^{UU}_{pq} \right)\nonumber \\
      &\hspace{2.5cm} - \mathbb{E}[\delta_1^i] \mathbb{E}[\beta_1^j] \nonumber
         \end{align}
    \begin{align}
       \mbox{cov}\left(\delta_{\frac{3}{2}}^i,\beta_{\frac{1}{2}}^j  \right) &= \left( \mathbb{E}[\mathbf{V}]\right)^{-1}_{ik}\left( \mathbb{E}[\mathbf{V}]\right)^{-1}_{jp} \left[ \mathbb{E}[{W}_{klm}] \left( \mathbb{E}[\mathbf{V}]\right)^{-1}_{lq}  \right. \nonumber \\
       &\hspace{1cm} \left( \left( \mathbb{E}[\mathbf{V}]\right)^{-1}_{mx} \left( \mathbb{E}[\mathbf{V}]\right)^{-1}_{yz} \Sigma^{VUUU}_{xyzpq} + {U}^\pi_x \Sigma^{UU}_{pq}\right) \nonumber \\
       &\hspace{0.5cm} + \frac{1}{2} \mathbb{E}[{W}_{klm}] \mathbb{E}[{W}_{xyz}] \left(\mathbb{E}[\mathbf{V}]\right)^{-1}_{lq}  \left(\mathbb{E}[\mathbf{V}]\right)^{-1}_{mx} \nonumber \\
       &\hspace{2cm}  \left(\mathbb{E}[\mathbf{V}]\right)^{-1}_{yu}  \left(\mathbb{E}[\mathbf{V}]\right)^{-1}_{zv} \Sigma^{UUUU}_{pquv} \nonumber \\
       &\hspace{0.5cm} + \frac{1}{2}  \left(\mathbb{E}[\mathbf{V}]\right)^{-1}_{lq}  \left(\mathbb{E}[\mathbf{V}]\right)^{-1}_{xr}  \left(\mathbb{E}[\mathbf{V}]\right)^{-1}_{ys} \nonumber \\
       &\hspace{2cm} \mathbb{E}[{W}_{qxy}] \Sigma^{VUUU}_{klrsp} \nonumber \\
       &\hspace{0.5cm} \left. +\frac{1}{2}  \left(\mathbb{E}[\mathbf{V}]\right)^{-1}_{ln}  \left(\mathbb{E}[\mathbf{V}]\right)^{-1}_{mq} \Sigma^{WUUU}_{klmnpq} \right]
              \end{align} 
       where
       \begin{align}
    \Sigma^{WUU}_{ijklm} &= \mathbb{E}\left[({W}_{ijk}-\mathbb{E}[{W}_{ijk}]){U}_{l} {U}_m\right] \nonumber \\
&= \frac{1}{n^2} \int l_{,ijk}(\mathbf{d})l_{,l}(\mathbf{d})l_{,m}(\mathbf{d})\; p(\mathbf{d} | \vec\lambda_t) {\rm d}\mathbf{d} \nonumber \\
& \hspace{2cm}-\mathbb{E}[{W}_{ijk}] \Sigma^{UU}_{lm} \nonumber \\
    \Sigma^{UUU}_{ijk} &= \mathbb{E}\left[{U}_{i}{U}_{j} {U}_k\right] \nonumber \\
&= \frac{1}{n^2} \int l_{,i}(\mathbf{d})l_{,j}(\mathbf{d})l_{,k}(\mathbf{d})\; p(\mathbf{d} | \vec\lambda_t) {\rm d}\mathbf{d} \nonumber \\
    \Sigma^{UUUU}_{ijkl} &= \mathbb{E}\left[{U}_{i} {U}_{j} {U}_{k} {U}_l\right] \nonumber \\
&= \frac{n(n-1)}{n^4} \left[ \int l_{,i}(\mathbf{d})l_{,j}(\mathbf{d})\; p(\mathbf{d} | \vec\lambda_t) {\rm d}\mathbf{d}  \right. \nonumber \\
&\hspace{2cm} \times \left.
 \int l_{,k}(\mathbf{d})l_{,l}(\mathbf{d})\; p(\mathbf{d} | \vec\lambda_t) {\rm d}\mathbf{d} \right]\nonumber \\
&\hspace{0.5cm} + \frac{1}{n^3} \int l_{,i}(\mathbf{d})l_{,j}(\mathbf{d})l_{,k}(\mathbf{d})l_{,l}(\mathbf{d})\; p(\mathbf{d} | \vec\lambda_t) {\rm d}\mathbf{d}  \nonumber \\
\Sigma^{VUUU}_{ijklm} &= \mathbb{E}\left[({V}_{ij}-\mathbb{E}[{V}_{ij}]){U}_k {U}_{l} {U}_m\right] \nonumber \\
&= \frac{n(n-1)}{n^4} \bigg[ \int l_{,ij}(\mathbf{d})l_{,k}(\mathbf{d})\; p(\mathbf{d} | \vec\lambda_t) {\rm d}\mathbf{d} \nonumber \\
&\hspace{1cm} \times \int l_{,l}(\mathbf{d})l_{,m}(\mathbf{d})\; p(\mathbf{d} | \vec\lambda_t) {\rm d}\mathbf{d} \nonumber \\
&\hspace{0.5cm}  + k \leftrightarrow l + k \leftrightarrow m \bigg]
%\int l_{,ij}(\mathbf{d})l_{,l}(\mathbf{d})\; p(\mathbf{d} | \vec\lambda_t) {\rm d}\mathbf{d}  \nonumber \\
%&\hspace{1cm} \times \int l_{,k}(\mathbf{d})l_{,m}(\mathbf{d})\; p(\mathbf{d} | \vec\lambda_t) {\rm d}\mathbf{d} \nonumber \\
%&\hspace{0.5cm} + \int l_{,ij}(\mathbf{d})l_{,m}(\mathbf{d})\; p(\mathbf{d} | \vec\lambda_t) {\rm d}\mathbf{d}  \nonumber \\
%&\hspace{1cm} \times \left. \int l_{,k}(\mathbf{d})l_{,l}(\mathbf{d})\; p(\mathbf{d} | \vec\lambda_t) {\rm d}\mathbf{d} \right]
\nonumber \\
&\hspace{0.2cm} + \frac{1}{n^3} \int l_{,ij}(\mathbf{d})l_{,k}(\mathbf{d})l_{,l}(\mathbf{d})l_{,m}(\mathbf{d})\; p(\mathbf{d} | \vec\lambda_t) {\rm d}\mathbf{d} \nonumber \\
&\hspace{2cm} -\mathbb{E}[{V}_{ij}] \Sigma^{UUU}_{klm} \nonumber \\
\Sigma^{WUUU}_{ijklmn} &= \mathbb{E}\left[({W}_{ijk}-\mathbb{E}[{W}_{ijk}]) {U}_{l} {U}_m {U}_n\right] \nonumber \\
&= \frac{n(n-1)}{n^4} \bigg[ \int l_{,ijk}(\mathbf{d})l_{,l}(\mathbf{d})\; p(\mathbf{d} | \vec\lambda_t) {\rm d}\mathbf{d} \nonumber \\
&\hspace{1cm} \times \int l_{,m}(\mathbf{d})l_{,n}(\mathbf{d})\; p(\mathbf{d} | \vec\lambda_t) {\rm d}\mathbf{d} \nonumber \\
&\hspace{0.5cm} + l \leftrightarrow m + l \leftrightarrow n \bigg]
%\int l_{,ijk}(\mathbf{d})l_{,m}(\mathbf{d})\; p(\mathbf{d} | \vec\lambda_t) {\rm d}\mathbf{d}  \nonumber \\
%&\hspace{1cm} \times \int l_{,l}(\mathbf{d})l_{,n}(\mathbf{d})\; p(\mathbf{d} | \vec\lambda_t) {\rm d}\mathbf{d} \nonumber \\
%&\hspace{0.5cm} + \int l_{,ijk}(\mathbf{d})l_{,n}(\mathbf{d})\; p(\mathbf{d} | \vec\lambda_t) {\rm d}\mathbf{d}  \nonumber \\
%&\hspace{1cm} \times \left. \int l_{,l}(\mathbf{d})l_{,m}(\mathbf{d})\; p(\mathbf{d} | \vec\lambda_t) {\rm d}\mathbf{d} \right]
\nonumber \\
&\hspace{0.2cm} + \frac{1}{n^3} \int l_{,ijk}(\mathbf{d})l_{,l}(\mathbf{d})l_{,m}(\mathbf{d})l_{,n}(\mathbf{d})\; p(\mathbf{d} | \vec\lambda_t) {\rm d}\mathbf{d} \nonumber \\ 
&\hspace{2cm} -\mathbb{E}[{W}_{ijk}] \Sigma^{UUU}_{lmn}.
\end{align}
From these we can construct the leading order covariances
\begin{align}
       \mbox{cov}\left( \beta^i, \delta^j\right) &= \mbox{cov}\left( \beta_{\frac{1}{2}}^i, \delta_1^j\right) + \mbox{cov}\left( \beta_1^i, \delta_1^j\right) + \mbox{cov}\left( \beta_{\frac{1}{2}}^i, \delta_{\frac{3}{2}}^j\right) \nonumber \\
       \mbox{cov}\left( \delta^i, \delta^j\right) &= \mbox{cov}\left( \delta_1^i, \delta_1^j\right).
\end{align}
Putting these together we can obtain the mean and covariance of the  posterior mode, $\hat{\lambda}^i-\lambda^i_t=\beta^i+\delta^i$, expressed as a deviation from the true population parameters.
\begin{align}
    \mathbb{E}[\hat{\lambda}^i-\lambda^i_t] &= \mathbb{E}[\beta^i] + \mathbb{E}[\delta^i] \nonumber \\
    \mbox{cov}\left(\hat{\lambda}^i-\lambda^i_t,\hat{\lambda}^j-\lambda^j_t \right) &=\mbox{cov}\left(\beta^i,\beta^j\right) + \mbox{cov}\left(\delta^i,\delta^j\right) \nonumber \\
    & \hspace{0.5cm}+ \mbox{cov}\left(\beta^i,\delta^j\right) +
    \mbox{cov}\left(\beta^j,\delta^i\right).
\end{align}
We now turn our attention to the posterior mean and variance. These are averages over the posterior. The definition of $\vec\beta$ was motivated to ensure the leading terms of the log-posterior can be written as a quadratic in $\vec\lambda - \vec\beta-\vec\lambda_t$. Denoting $\vec B=\vec\lambda_t+\vec\beta$, averages of a function $f(\vec\lambda)$ over the posterior then take the form
\begin{align}
\langle f \rangle &=\frac{\int f(\vec\lambda)  g(\vec\lambda)  \exp\left[ -\frac{n}{2} (\lambda^i-B^i) (V_{ij} + V^\pi_{ij})(\lambda^j-B^j) \right] {\rm d}\vec\lambda}{{\int g(\vec\lambda)  \exp\left[ -\frac{n}{2} (\lambda^i-B^i) (V_{ij} + V^\pi_{ij})(\lambda^j-B^j) \right] {\rm d}\vec\lambda}} \nonumber \\
\ln g(\vec\lambda) &= -\frac{n}{6} W_{ijk} (\lambda^i-\lambda_t^i) (\lambda^j-\lambda_t^j) (\lambda^k-\lambda_t^k) \nonumber \\
&\hspace{0.5cm} - \frac{n}{24} X_{ijkl} (\lambda^i-\lambda_t^i) (\lambda^j-\lambda_t^j) (\lambda^k-\lambda_t^k) (\lambda^l-\lambda_t^l) \nonumber \\
X_{ijkl} &= \left( \frac{\partial^4 \hat\mu}{\partial\lambda^i \partial\lambda^j \partial\lambda^k \partial\lambda^l}\right)_{|\vec\lambda_t}
\end{align}
There are further corrections in $g(\vec\lambda)$ from higher derivates in the expansion, and from the prior terms, $W_{ijk}^\pi$, $X_{ijkl}^\pi$ etc. However, the contributions from the included terms can be seen to be $1/n$ down relative to the leading terms in the integral, and these other corrections are at least $1/n^{\frac{3}{2}}$ down from leading and are hence sub-dominant.

Integrals of these form are standard and we will make use of the following results
 \begin{align}\label{eq:standard_integrals}
I_0(\Gamma)&= \int \exp\left[-\frac{1}{2} \mathbf{x}^T \Gamma^{-1} \mathbf{x} \right] {\rm d}\mathbf{x} = (2\pi)^\frac{N}{2} \sqrt{|\Gamma|} \\ 
I_{ij}(\Gamma)&= \int x_i x_j \exp\left[-\frac{1}{2} \mathbf{x}^T \Gamma^{-1} \mathbf{x} \right] {\rm d}\mathbf{x} \nonumber \\
&=(2\pi)^\frac{N}{2} \sqrt{|\Gamma|} \Gamma_{ij} \\
 I_{ijkl}(\Gamma)&=\int x_i x_j x_k x_l\exp\left[-\frac{1}{2} \mathbf{x}^T \Gamma^{-1} \mathbf{x} \right] {\rm d}\mathbf{x} \nonumber \\ &=(2\pi)^\frac{N}{2} \sqrt{|\Gamma|} \left(\Gamma_{ij} \Gamma_{kl} + \Gamma_{ik}\Gamma_{jl} + \Gamma_{il}\Gamma_{jk} \right) \\
I_{ijklmn}(\Gamma)&= \int x_i x_j x_k x_l x_m x_n \exp\left[-\frac{1}{2} \mathbf{x}^T \Gamma^{-1} \mathbf{x} \right] {\rm d}\mathbf{x} \nonumber \\
&=(2\pi)^\frac{N}{2} \sqrt{|\Gamma|} \left(\Gamma_{ij}\Gamma_{kl}\Gamma_{mn} + \Gamma_{ij}\Gamma_{km}\Gamma_{ln}  \right. \nonumber \\
 & \hspace{0.5cm}  + \Gamma_{ij}\Gamma_{kn}\Gamma_{lm} 
+\Gamma_{ik}\Gamma_{jl}\Gamma_{mn} +\Gamma_{ik}\Gamma_{jm}\Gamma_{ln} \nonumber \\
 & \hspace{0.5cm}+\Gamma_{ik}\Gamma_{jn}\Gamma_{lm} 
+\Gamma_{il}\Gamma_{jk}\Gamma_{mn} +\Gamma_{il}\Gamma_{jm}\Gamma_{kn}  \nonumber \\
 & \hspace{0.5cm} +\Gamma_{il}\Gamma_{jn}\Gamma_{km} +\Gamma_{im}\Gamma_{jk}\Gamma_{ln} +\Gamma_{im}\Gamma_{jl}\Gamma_{kn}  \nonumber \\
 & \hspace{0.5cm}+\Gamma_{im}\Gamma_{jn}\Gamma_{kl} +\Gamma_{in}\Gamma_{jk}\Gamma_{lm} +\Gamma_{in}\Gamma_{jl}\Gamma_{km}  \nonumber \\
 &\hspace{1cm} \left.+\Gamma_{in}\Gamma_{jm}\Gamma_{kl} \right)
  \end{align}
where $|\Gamma|$ denotes the determinant of $\Gamma$. We will also use the notation $\tilde{I}_{ij}(\Gamma) \equiv I_{ij}(\Gamma)/I_0(\Gamma)$ and similarly for other terms. In this case, the covariance matrix $\Gamma=(\mathbf{V}+\mathbf{V}^\pi)^{-1}/n$. Every additional factor of $\Gamma$ therefore introduces an extra negative power of $n$. This allows us to identify the dominant terms. To evaluate the above expressions we need to be able to compute $\Gamma$, which can be done perturbatively by noting 
\begin{equation}
n\Gamma(\mathbb{E}[\mathbf{V}] + (\mathbf{V}-\mathbb{E}[\mathbf{V}]) + \mathbf{V}^\pi) = \mathbf{I}
\end{equation}
from which
\begin{align}
\Gamma &= \frac{1}{n} \left(\Gamma_0 + \Gamma_{\frac{1}{2}} + \Gamma_{1} + \cdots \right) \nonumber \\
\Gamma_0 &= (\mathbb{E}[\mathbf{V}])^{-1} \nonumber \\
\Gamma_{\frac{1}{2}} &= - (\mathbb{E}[\mathbf{V}])^{-1} \left( \mathbf{V} - \mathbb{E}[\mathbf{V}]\right) \nonumber \\
\Gamma_{1} &=-(\mathbb{E}[\mathbf{V}])^{-1} \left[ \Gamma_{\frac{1}{2}} \left( \mathbf{V} - \mathbb{E}[\mathbf{V}]\right) + \mathbf{V}^\pi \right].
\end{align}
To obtain the posterior mean, we first compute
\begin{align}
\langle g \rangle&=\int g(\vec\lambda)  \exp\left[ -\frac{n}{2} (\lambda^i-B^i) (V_{ij} + V^\pi_{ij})(\lambda^j-B^j) \right] {\rm d}\vec\lambda \nonumber \\
&= I_0(\Gamma) - \frac{n}{6} W_{ijk} \left[ \beta^i I_{jk}(\Gamma) + \beta^j I_{kl}(\Gamma) + \beta^l I_{ij}(\Gamma) \right] \nonumber \\
&\hspace{0.5cm} - \frac{n}{6} W_{ijk}
\beta^i \beta^j \beta^k I_0(\Gamma) + \frac{n^2}{72} W_{ijk} W_{lmn} I_{ijklmn} (\Gamma) \nonumber \\
&\hspace{2cm} -\frac{n}{24} X_{ijkl} I_{ijkl}(\Gamma).
\end{align}
Using similar notation to before we can write
\begin{align}
\langle g \rangle&=I_0(\Gamma) \left( 1 + g_{\frac{1}{2}} + g_1 + \cdots \right) \nonumber\\
g_{\frac{1}{2}}&= -\frac{1}{6} \mathbb{E}[{W}_{ijk}] \left( \beta^i_{\frac{1}{2}} (\Gamma_0)_{jk} + \beta^j_{\frac{1}{2}} (\Gamma_0)_{kl} + \beta^l_{\frac{1}{2}} (\Gamma_0)_{ij}\right) \nonumber \\
&\hspace{1cm} -\frac{n}{6} \mathbb{E}[{W}_{ijk}] \beta^i_{\frac{1}{2}} \beta^j_{\frac{1}{2}} \beta^k_{\frac{1}{2}} \nonumber \\
g_1 &=-\frac{1}{6} ({W}_{ijk}-\mathbb{E}[{W}_{ijk}]) \nonumber \\
&\hspace{1cm} \times \left( \beta^i_{\frac{1}{2}} (\Gamma_0)_{jk} + \beta^j_{\frac{1}{2}} (\Gamma_0)_{kl} + \beta^l_{\frac{1}{2}} (\Gamma_0)_{ij})\right) \nonumber \\
&\hspace{0.5cm} -\frac{1}{6} \mathbb{E}[{W}_{ijk}] \left( \beta^i_{\frac{1}{2}} (\Gamma_{\frac{1}{2}})_{jk} + \beta^j_{\frac{1}{2}} (\Gamma_{\frac{1}{2}})_{kl} + \beta^l_{\frac{1}{2}} (\Gamma_{\frac{1}{2}})_{ij} \right. \nonumber \\
&\hspace{1cm} \left. + \beta^i_{1} (\Gamma_0)_{jk} + \beta^j_{1} (\Gamma_0)_{kl} + \beta^l_{1} (\Gamma_0)_{ij} \right) \nonumber \\
&\hspace{0.5cm} - \frac{n}{6} ({W}_{ijk}-\mathbb{E}[{W}_{ijk}]) \beta^i_{\frac{1}{2}} \beta^j_{\frac{1}{2}} \beta^k_{\frac{1}{2}} \nonumber \\
&\hspace{0.5cm} - \frac{n}{6} \mathbb{E}[{W}_{ijk}] \left( \beta^i_{1} \beta^j_{\frac{1}{2}} \beta^k_{\frac{1}{2}} + \beta^i_{\frac{1}{2}} \beta^j_{1} \beta^k_{\frac{1}{2}} + \beta^i_{\frac{1}{2}} \beta^j_{\frac{1}{2}} \beta^k_{1} \right) \nonumber \\
&\hspace{0.5cm} + \frac{1}{72}  \mathbb{E}[{W}_{ijk}]  \mathbb{E}[{W}_{lmn}] \frac{1}{n} \tilde{I}_{ijklmn}(\Gamma_0) \nonumber \\
&\hspace{0.5cm}- \frac{1}{24} \frac{1}{n}  \mathbb{E}[{X}_{ijkl}] \tilde{I}_{ijkl}(\Gamma_0)
\end{align}

Now we compute the posterior mean, expressed as a distance from the true population parameters
\begin{align}
\bar\lambda^i - \lambda^i_t &\equiv \langle (\lambda^i - \lambda^i_t) \rangle = \beta^i +  \langle (\lambda^i - B^i) \rangle\nonumber \\
&=\beta^i + \frac{1}{\langle g \rangle} \left[ -\frac{n}{6} W_{jkl} \left(I_{ijkl}(\Gamma) + 3 \beta^j \beta^k I_{il}(\Gamma) \right)\right. \nonumber \\
%&\hspace{1cm} \left. + \beta^j \beta^l I_{ik}(\Gamma) + \beta^k \beta^ll I_{ij}(\Gamma) \right) \nonumber \\
&\hspace{1cm} -\frac{n}{6} X_{jklm} \left( \beta^j I_{iklm}(\Gamma) + \beta^j\beta^k\beta^l I_{im}(\Gamma) \right) \nonumber \\
&\hspace{1cm} +\frac{n^2}{36} W_{jkl} W_{mnp} \left( 3 \beta^j I_{iklmnp}(\Gamma) 
\right. \nonumber \\
&\hspace{2cm} + \beta^j\beta^k\beta^l I_{imnp}(\Gamma) + 3 \beta^j\beta^k\beta^m I_{ilnp} (\Gamma) \nonumber \\
&\hspace{2cm} \left.+ 3\beta^j\beta^k\beta^l \beta^m\beta^n I_{ip}(\Gamma)\right) %may have miscounted terms here
%+  \beta^k I_{ijlm}(\Gamma) +  \beta^l I_{ijkm}(\Gamma) +  \beta^m I_{ikjl}(\Gamma) \right)
\left. \right].
\end{align}
We note that the leading order correction in the bracketed term is $1/n$. To obtain the posterior mean to the same order as $\beta$ we therefore only need to retain terms up to $g_{\frac{1}{2}}$ in $\langle g \rangle$. Specifically we can write
\begin{align}
    \bar\lambda^i - \lambda^i_t &=\bar\lambda^i_{\frac{1}{2}} + \bar\lambda^i_{1} + \bar\lambda_{\frac{3}{2}} + \cdots \nonumber \\
    \bar\lambda^i_{\frac{1}{2}} &= \beta^i_{\frac{1}{2}} \nonumber \\
    \bar\lambda^i_{1} &= \beta^i_{1} -\frac{1}{6} \mathbb{E}[{W}_{jkl}] \left( \frac{1}{n} \tilde{I}_{ijkl}(\Gamma_0) + 3\beta_{\frac{1}{2}}^j \beta_{\frac{1}{2}}^k (\Gamma_0)_{il}\right) \nonumber \\
    \bar\lambda^i_{\frac{3}{2}} &= \beta^i_{\frac{3}{2}} + \left( \frac{g_{\frac{1}{2}}}{6} \mathbb{E}[{W}_{jkl}]  -\frac{1}{6} (W_{jkl}-\mathbb{E}[{W}_{jkl}])\right) \nonumber \\
    &\hspace{2cm} \times \left( \frac{1}{n} \tilde{I}_{ijkl}(\Gamma_0) + 3\beta_{\frac{1}{2}}^j \beta_{\frac{1}{2}}^k (\Gamma_0)_{il}\right)\nonumber \\
    &\hspace{0.5cm} -\frac{1}{6} \mathbb{E}[{W}_{jkl}] \left( \frac{2}{n} \left((\Gamma_0)_{ij} (\Gamma_{\frac{1}{2}})_{kl} + (\Gamma_0)_{ik} (\Gamma_{\frac{1}{2}})_{jl} \right. \right . \nonumber \\
    &\hspace{3cm} \left. + (\Gamma_0)_{il} (\Gamma_{\frac{1}{2}})_{jk} \right) \nonumber \\
    &\hspace{2cm} + 3(\beta_1^j \beta^k_{\frac{1}{2}} + \beta^j_{\frac{1}{2}} \beta^k_1) (\Gamma_0)_{il} \nonumber \\
    &\hspace{2.5cm} \left. + 3 \beta^j_{\frac{1}{2}} \beta^k_{\frac{1}{2}} (\Gamma_{\frac{1}{2}})_{il} \right) \nonumber \\
    &\hspace{0.5cm} - \frac{1}{6} \mathbb{E}[X_{jklm}] \left( \beta^j_{\frac{1}{2}} \frac{1}{n} \tilde{I}_{iklm} + \beta^j_{\frac{1}{2}} \beta^k_{\frac{1}{2}} \beta^l_{\frac{1}{2}} (\Gamma_0)_{im}\right) \nonumber \\
    &\hspace{0.5cm} +\frac{1}{36} \mathbb{E}[W_{jkl}] \mathbb{E}[W_{mnp}] \left( \frac{3}{n}\beta^j_{\frac{1}{2}} \tilde{I}_{iklmnp}(\Gamma_0) \right. \nonumber \\
    &\hspace{1cm} + \beta^j_{\frac{1}{2}} \beta^k_{\frac{1}{2}} \beta^l_{\frac{1}{2}} \tilde{I}_{imnp} (\Gamma_0) + 3 \beta^j_{\frac{1}{2}} \beta^k_{\frac{1}{2}} \beta^m_{\frac{1}{2}} \tilde{I}_{ilnp}(\Gamma_0) \nonumber \\
    &\hspace{2cm} \left. + 3 n \beta^j_{\frac{1}{2}} \beta^k_{\frac{1}{2}} \beta^l_{\frac{1}{2}} \beta^m_{\frac{1}{2}} \beta^n_{\frac{1}{2}} (\Gamma_0)_{ip} \right)
\end{align}
Finally, we consider the posterior covariance
 \begin{align}
 \hat\Gamma_{ij} &\equiv \langle (\lambda^i - \bar\lambda^i) (\lambda^j - \bar\lambda^j) \rangle \nonumber \\
 &= \langle (\lambda^i - B^i) (\lambda^j - B^j) \rangle \nonumber \\
 &\hspace{1cm} + (B^i - \bar\lambda^i) \langle  (\lambda^j - B^j) \rangle + i \leftrightarrow j \nonumber \\
 &\hspace{2cm}+ (B^i - \bar\lambda^i) (B^j - \bar\lambda^j).
 \end{align}
The term on the first line has a leading order dependence of $1/n$, plus corrections of $1/n^{\frac{3}{2}}$. The terms on the second and third lines are $O(1/n^2)$ and so are sub-dominant. We deduce
\begin{align}
 \hat\Gamma_{ij} &= \frac{1}{\langle g \rangle} \left[I_{ij}(\Gamma) -\frac{n}{6} W_{klm} \left(3\beta^k I_{ijlm}(\Gamma) + \beta^k\beta^l\beta^m I_{ij}(\Gamma)  \right) \right]
\end{align}
and expand
\begin{align}
    \hat\Gamma_{ij} &= (\hat\Gamma_1)_{ij} + \left(\hat\Gamma_\frac{3}{2}\right)_{ij} + \cdots \nonumber \\
    (\hat\Gamma_1)_{ij} &= \frac{1}{n} \left( \Gamma_0 \right)_{ij} \nonumber \\
    \left(\hat\Gamma_\frac{3}{2}\right)_{ij} &= \frac{1}{n} \left( \Gamma_{\frac{1}{2}} \right)_{ij} - \frac{g_{\frac{1}{2}}}{n} \left( \Gamma_0 \right)_{ij} \nonumber \\
    &\hspace{0.5cm} - \frac{1}{6} \mathbb{E}[W_{klm}] \left( \frac{3}{n} \beta_{\frac{1}{2}}^k \tilde{I}_{ijlm}(\Gamma_0) + \beta_{\frac{1}{2}}^k \beta_{\frac{1}{2}}^l \beta_{\frac{1}{2}}^m (\Gamma_0)_{ij}\right)
\end{align}
 Using the preceding expressions, we could now compute the mean and variance of the posterior mean and covariance as we did for the shift in the posterior mode. However, this calculation is very similar to the calculations carried out above and is tedious so we leave it out. Instead we note a number of features.
 \begin{itemize}
 \item The leading order posterior covariance is constant and equal to $(\mathbb{E}[\mathbf{V}])^{-1} /n$, which is the expression we used to derive the population Fisher matrix in the body of the paper.
 \item The leading order difference between either the posterior mode or mean and the true parameter value has expectation value that scales like $1/n$ and variance that also scales like $1/n$. Thus, the posterior bias is noise dominated, i.e., fluctuations due to the particular random realisation of the population that was observed dominate over the fixed bias. If multiple sets of observations of $n$ events were repeated and averaged, then this bias would eventually be significant. In practice we would never do this since it weakens the precision of inference. This means that computing corrections to the posterior mean is unnecessary.
 \item Similarly, the expectation value of $\left(\hat\Gamma_\frac{3}{2}\right)_{ij}$ is zero, so the leading order correction to the expected value of the posterior covariance scales like $n^{-2}$, while the variance in the posterior covariance scales like $n^{-3}$. So, fluctuations in the posterior covariance due to randomness in the observed population are larger than the size of corrections from the finite number of observations.
 \end{itemize}

\section{Generalisation to other likelihoods}
\label{app:GenLike}
Expression~\ref{eq:gammalambda} was derived for the GW likelihood defined by Eq.~(\ref{eq:innprod}), but it can be extended to more general likelihoods, $p(\mathbf{d}|\vec\theta)$. The likelihood enters the result through the definitions of $\Gamma$, $N_i$, $D_i$ and $D_{ij}$. For a more general likelihood we have
\begin{align}
    \Gamma_{ij} &= -\frac{\partial^2 \ln p(\mathbf{d}|\vec\theta)}{\partial\theta^i \partial \theta^j} \nonumber \\
    N_i &= \frac{\partial \ln p(\mathbf{d}|\vec\theta)}{\partial\theta^i},
\end{align}
where derivatives are evaluated at $\vec\theta_0$. In the gravitational wave case these become
\begin{align}
    \Gamma_{ij} &= \left(\frac{\partial \mathbf{h}}{\partial\theta^i} \bigg| \frac{\partial \mathbf{h}}{\partial\theta^j} \right) -\left( \mathbf{d} - \mathbf{h}(\vec\theta_0) \bigg| \frac{\partial^2 \mathbf{h}}{\partial\theta^i \partial \theta^j}\right) \nonumber \\
    N_i &= \left( \mathbf{d} - \mathbf{h}(\vec\theta_0) \bigg| \frac{\partial^2 \mathbf{h}}{\partial\theta^i \partial \theta^j}\right).
\end{align}
Dropping the second term in the expressions for $\Gamma$ on the grounds that it is smaller by a factor of $\rho^{-1}$ than the first, we recover the expressions used in the earlier derivation. In particular, we note that with this simplification $\Gamma_{ij}$ does not depend on $\mathbf{d}$ and hence we can take the terms that depend on $\Gamma$ outside of the integral over data, simplifying the final form of the population Fisher matrix. For a more general likelihood we can not assume this is the case, but have $\Gamma_{ij}(\mathbf{d},\vec\theta_0)$ and $N_i(\mathbf{d},\vec\theta_0)$. The first contributions to the population likelihood, $\Gamma_\text{I}$ takes the same form as before, but the other contributions are modified to
\begin{align}
(\Gamma_\text{II})_{ij}&= \frac{1}{2} \int\int \frac{\partial^2 \ln {\rm det}(\Gamma+H)}{\partial\lambda^i \partial\lambda^j} \nonumber \\
&\hspace{2cm} \times p(\mathbf{d} | \vec\theta_0 ) \frac{p(\vec\theta_0 | \vec\lambda)}{P_{\rm det}(\vec\lambda)} {\rm d} \mathbf{d} {\rm d} \vec\theta_0,\nonumber \\
(\Gamma_\text{III})_{ij}&= -\frac{1}{2} \int\int \frac{\partial^2}{\partial\lambda^i \partial\lambda^j}\left[(\Gamma+H)^{-1}_{kl}\right] N_{k} N_{l} \nonumber \\
&\hspace{2cm} \times p(\mathbf{d} | \vec\theta_0 ) \frac{p(\vec\theta_0 | \vec\lambda)}{P_{\rm det}(\vec\lambda)} {\rm d} \mathbf{d} {\rm d} \vec\theta_0, \nonumber \\
(\Gamma_\text{IV})_{ij}&=  -\int\int \frac{\partial^2}{\partial\lambda^i \partial\lambda^j} \left[ P_k(\Gamma+H)^{-1}_{kl}\right]N_{l}  \nonumber \\
&\hspace{2cm} \times  p(\mathbf{d} | \vec\theta_0 ) \frac{p(\vec\theta_0 | \vec\lambda)}{P_{\rm det}(\vec\lambda)} {\rm d} \mathbf{d} {\rm d} \vec\theta_0, \nonumber \\
(\Gamma_\text{V})_{ij}&= -\frac{1}{2} \int\int \frac{\partial^2}{\partial\lambda^i \partial\lambda^j} \left[ P_k (\Gamma+H)^{-1}_{kl} P_l \right]  \nonumber \\
&\hspace{2cm} \times p(\mathbf{d} | \vec\theta_0 ) \frac{p(\vec\theta_0 | \vec\lambda)}{P_{\rm det}(\vec\lambda)} {\rm d} \mathbf{d} {\rm d} \vec\theta_0\nonumber,
\end{align}
where the integrals over $\mathbf{d}$ are over detectable data sets. A further simplification can be obtained if we assume that the variance of $\Gamma_{ij}$ over realisations of the data, $p(\mathbf{d}|\vec\theta_0)$, is small. This allows us to use the fact that the expectation value of any function, $f(X,Y)$, of two random variables $X$ and $Y$, can be expanded
\begin{align}
    \mathbb{E}\left[f(X,Y) \right] &= f(\mathbb{E}[X],\mathbb{E}[Y]) + \frac{1}{2} {\rm Var}(X) \left(\frac{\partial^2 f}{\partial X^2}\right)_{(\mathbb{E}(X),\mathbb{E}(Y))} \nonumber \\
    &\hspace{1cm} + {\rm Cov}(X,Y) \left(\frac{\partial^2 f}{\partial X \partial Y}\right)_{(\mathbb{E}(X),\mathbb{E}(Y))} \nonumber \\
    &\hspace{1cm} + \frac{1}{2} {\rm Var}(Y) \left(\frac{\partial^2 f}{\partial Y^2}\right)_{(\mathbb{E}(X),\mathbb{E}(Y))} + \cdots
\end{align}
Ignoring all but the leading term allows us to replace $\Gamma$ by its expectation value. With this additional assumption, the population Fisher matrix for the general case takes the sane form as before, with the substitutions $\Gamma_{ij} \rightarrow \bar{\Gamma}_{ij}$, $D_{i} \rightarrow \bar{D}_{i}$ and $D_{ij} \rightarrow \bar{D}_{ij}$, where
\begin{align}
\bar{\Gamma}_{ij} &= \int \left[- \frac{\partial^2 \ln p(\mathbf{d}|\vec\theta)}{\partial\theta^i \partial \theta^j}\right] p(\mathbf{d} | \vec\theta_0 ) {\rm d} \mathbf{d} \nonumber \\
\bar{D}_i &= \int \left[\frac{\partial \ln p(\mathbf{d}|\vec\theta)}{\partial\theta^i}\right] p(\mathbf{d} | \vec\theta_0 ) {\rm d} \mathbf{d} = \frac{\partial P_{\rm det}(\vec\theta)}{\partial \theta^i}\nonumber \\
\bar{D}_{ij} &= \int \left[ \frac{\partial \ln p(\mathbf{d}|\vec\theta)}{\partial\theta^i} \frac{\partial \ln p(\mathbf{d}|\vec\theta)}{\partial\theta^j}\right] p(\mathbf{d} | \vec\theta_0 ) {\rm d} \mathbf{d}. \nonumber \\
\end{align}

\section{Calculation of the Fisher Matrix for the power-law population with SNR distribution}
\label{app:GWlike_real_FM}
Here we provide a guide to computing the Fisher matrix for the more realistic GW-like example described in Section~\ref{sec:GWlike_real}. The source parameters are $\vec\theta = (\rho, M)$ and the population parameter is $\vec\lambda=(\alpha)$.
 
The full population model is
\begin{align}
    p(\vec\theta | \vec\lambda) &= \frac{3 M^3}{d_{\rm max}^3} \frac{1}{\rho^4} \frac{\alpha M^{\alpha-1}}{M_{\rm max}^\alpha - M_{\rm min}^\alpha} \nonumber \\
    &= \frac{3\alpha}{d_{\rm max}^3 (M_{\rm max}^\alpha - M_{\rm min}^\alpha)} \frac{1}{\rho^4} M^{2+\alpha}
\end{align}
from which we can deduce
\begin{equation}
    H_{ij} = \left( \begin{array}{cc} -\frac{4}{\rho^2}&0\\0&\frac{(2+\alpha)}{M^2} \end{array}\right).
\end{equation}
The single event Fisher matrix, $\Gamma$, was given in Eq.~(\ref{eq:gamma_theta_GWlike}). The detection probability is
\begin{equation}
P_{\rm det}(\vec\theta) = \frac{1}{2} {\rm erfc}\left[-\frac{(\rho_{\rm th}-\rho)}{\sqrt{2}}\right].
\end{equation}

The determinant of $\Gamma + H$ is
\begin{equation}
    {\rm det}(\Gamma + H) =\frac{rho^2}{\sigma_M^2}-\frac{4}{
    \sigma_M^2}+\frac{(2+\alpha)}{M^2} -\frac{4(2+\alpha)}{\rho^2M^2}+\frac{(2+\alpha)}{\sigma_M^2}
\end{equation}
which has first derivative
\begin{equation}
    \frac{\partial}{\partial\alpha} {\rm det}(\Gamma + H) = \frac{1}{M^2}+\frac{1}{\sigma_M^2}-\frac{4}{
    \rho^2M^2}
\end{equation}
and the second derivative vanishes. The integrals required for the matrices $\Gamma_{III}$, $\Gamma_{IV}$ and $\Gamma_V$ all take the form
\begin{equation}
    \Gamma_X = -\int \frac{\partial^2}{\partial\alpha^2} \left(\frac{A_X}{{\rm det}(\Gamma+H)}\right) \frac{p(\vec\theta_0|\alpha)}{P_{\rm det}(\alpha)} {\rm d}\vec\theta_0,
\end{equation}
where
\begin{align}
    A_{III} &=\frac{1}{2}\left(\frac{\rho^2}{\sigma_M^2}+\frac{(2+\alpha)}{M^2} \right) \frac{(\rho_{\rm th}-\rho)}{\sqrt{2\pi}} \exp\left[-\frac{(\rho_{\rm th}-\rho)^2}{2}\right] \nonumber \\
    &\hspace{1cm} + \left(\frac{\rho^2}{\sigma_M^2}-\frac{(2+\alpha)}{2\sigma_M^2}+\frac{(2+\alpha)}{2M^2} \right) P_{\rm det}(\theta_0) \nonumber \\
    A_{IV} &= -\left(\frac{(6+\alpha)\rho}{\sigma_M^2}+\frac{4(2+\alpha)}{\rho M^2} \right) \frac{1}{\sqrt{2\pi}} \exp\left[-\frac{(\rho_{\rm th}-\rho)^2}{2}\right]\nonumber \\
    A_{V} &= \left(\frac{8}{\sigma_M^2}-\frac{8(2+\alpha)}{\rho^2M^2}+\frac{4(2+\alpha)}{\sigma_M^2}-\frac{2(a+\alpha)^2}{\rho^2M^2} \right. \nonumber \\
    &\hspace{1cm} \left. + \frac{(2+\alpha)^2}{2M^2} +  \frac{(2+\alpha^)2}{2 \sigma_M^2} \right) P_{\rm det}(\vec\theta_0).
\end{align}
These terms have first derivatives
\begin{align}
    \frac{\partial A_{III}}{\partial\alpha} &= \frac{(\rho_{\rm th}-\rho)}{\sqrt{2\pi} M^2} \exp\left[-\frac{(\rho_{\rm th}-\rho)^2}{2}\right] \nonumber \\
    &\hspace{1cm} + \frac{1}{2} \left(\frac{1}{\sigma_M^2} + \frac{1}{M^2}\right) P_{\rm det}(\vec\theta_0) \\
    \frac{\partial A_{IV}}{\partial\alpha} &= -\left(\frac{4}{\rho^2M^2}+\frac{\rho}{\sigma_M^2} \right) \frac{1}{\sqrt{2\pi}} \exp\left[-\frac{(\rho_{\rm th}-\rho)^2}{2}\right] \nonumber \\
    \frac{\partial A_{V}}{\partial\alpha} &=\left(\frac{(2+\alpha)}{M^2}+\frac{(6+\alpha)}{\sigma_M^2}-\frac{4\alpha}{\rho^2M^2}\right)P_{\rm det}(\vec\theta_0)
\end{align}
and all second derivatives vanish except
\begin{align}
    \frac{\partial^2 A_{V}}{\partial\alpha^2} &=\left(-\frac{4}{\rho^2M^2}+\frac{1}{M^2}+\frac{1}{\sigma_M^2}\right) P_{\rm det}(\vec\theta_0).
\end{align}
These expressions allow all of the integrands that determine the different parts of the population Fisher matrix to be valuated. The final stage of computing the Fisher matrix is to carry out the integrals over the population distribution $p(\vec\theta_0|\vec\lambda)$. This must be done numerically, but it is facilitated by doing a coordinate transformation
\begin{align}
    v(\vec\theta_0) &= \frac{M^\alpha - M_{\rm min}^\alpha}{M_{\rm max}^\alpha - M_{\rm min}^\alpha} \\
    u(\vec\theta_0) &= 1-\left(\frac{M}{d_{\rm max}\rho} \right)^3,
\end{align}
which reduces the population integral
\begin{equation}
    \int p(\vec\theta_0|\lambda) {\rm d}\vec\theta_0 \rightarrow \int {\rm d}u {\rm d}v.
\end{equation}
Further computational efficiencies can be obtained by restricting the range of $u$ considered for each $v$ so that only SNRs $\rho > \rho_{\rm th} - 5$ are included. Codes to compute the Fisher matrix using this procedure are available at \url{https://github.com/aantonelli94/PopFisher}.

\end{document}